%
%
%

\baselineskip 14pt plus 2pt

\font\llbf=cmbx10 scaled\magstep2
\font\lbf=cmbx10 scaled\magstep1

\def\ni{\noindent}

\def\ua{\underline a \,}
\def\ub{\underline b \,}
\def\uc{\underline c \,}
\def\ud{\underline d \,}

\def\bi{\bf i \,}
\def\bj{\bf j \,}
\def\bk{\bf k \,}
\def\bl{\bf l \,}
\def\bm{\bf m \,}
\def\bn{\bf n \,}


\ni
{\llbf On the roots of the Poincare structure of asymptotically flat 
spacetimes}\footnote{${}^*$}{Dedicated to Jim Nester on the occasion 
of his 60th birthday.}\par
\bigskip
\bigskip
\ni
{\bf L\'aszl\'o B. Szabados}\par
\bigskip
\ni
Research Institute for Particle and Nuclear Physics \par
\ni
H-1525 Budapest 114, P.O.Box 49, Hungary \par
\ni
E-mail: lbszab@rmki.kfki.hu \par
\bigskip
\bigskip

\ni
The analysis of canonical vacuum general relativity by R. Beig and 
N. \'O Murchadha (Ann. Phys. {\bf 174} 463-498 (1987)) is extended 
in numerous ways. The weakest possible power-type fall-off 
conditions for the energy-momentum tensor of the matter fields, 
the metric, the extrinsic curvature, the lapse and the shift are 
determined which, together with the parity conditions, are 
preserved by the energy-momentum conservation law $T^{ab}{}_{;b}=0$ 
and the evolution equations for the geometry. The algebra of the 
asymptotic Killing vectors, defined with respect to a foliation of 
the spacetime, is shown to be the Lorentz Lie algebra for slow 
fall-off of the metric, but it is the Poincare algebra for $1/r$ or 
faster fall-off. 

It is shown that the applicability of the symplectic formalism 
already requires the $1/r$ (or faster) fall-off of the metric. The 
connection between the Poisson algebra of the Beig--\'O Murchadha 
Hamiltonians (and, in particular, the constraint algebra) and the 
asymptotic Killing vectors is clarified. Their Hamiltonian $H[K^a]$ 
is shown to be constant in time modulo constraints for those 
asymptotic Killing vectors $K^a$ that are defined with respect to 
the foliation by the constant time slices. 

The energy-momentum and angular momentum are defined by the boundary 
term $Q[K^a]$ in $H[K^a]$ even in the presence of matter. Although 
the energy-momentum is well defined even for slightly faster than 
the $r^{-1/2}$ fall-off, we show that the angular momentum and 
centre-of-mass are finite only if the metric falls off as $1/r$ or 
faster. $Q[K^a]$ is constant in time for those $K^a$'s that are 
asymptotic Killing vectors with respect to the foliation by the 
constant time slices. 
If the foliation corresponds to proper time evolution (i.e. its lapse 
tends to 1 at infinity), then $Q[K^a]$ reproduces the ADM energy, the 
spatial momentum and spatial angular momentum, but the centre-of-mass 
deviates from that of Beig and \'O Murchadha by the spatial momentum 
times the coordinate time. The spatial angular momentum and the new 
centre-of-mass form an anti-symmetric Lorentz tensor, which 
transforms in the expected way under Poincare transformations. 

\bigskip
\bigskip
\vfill\eject
\ni
{\lbf 1. Introduction}\par
\medskip
\ni
The quantum field theoretical investigations of the early sixties 
showed that, strictly speaking, the observables of quantum fields 
must be associated only with finite but extended spacetime domains, 
i.e. they are {\it quasi-local} [1]. Quantities associated with 
spacetime points are not observables, and the global quantities, 
e.g. the total energy or electric charge, should be considered as 
the limit of quasi-locally defined quantities. 
Interestingly enough (although by different reasons, but) the 
situation is very similar in general relativity: energy-momentum 
and angular momentum cannot be associated with the points of the 
spacetime. Any such local expression is necessarily pseudotensorial 
and/or internal gauge dependent. Thus if we want to characterize the 
gravitational `field' by observables finer than those associated 
with the {\it whole} (necessarily asymptotically flat) spacetime, 
then these observables must also be defined quasi-locally. 

In the last two decades a lot of efforts was concentrated on the 
investigations both of the general framework in which the quasi-local 
energy-momentum and angular momentum should be constructed and the 
specific constructions themselves (and their properties). Although 
there is no consensus at all in the relativity community even about 
general questions e.g. when to consider a specific construction to 
be `reasonable', it is naturally expected that the globally defined 
observables, e.g. the ADM energy-momentum, must be recoverable as 
an appropriate limit of the corresponding quasi-local quantities. 
Although the energy-momentum, both at the spatial and null infinity, 
is well understood, the (relativistic) angular momentum (especially 
at the null infinity) needs more investigations. In particular, one 
should clarify the limit of the spatial angular momentum and 
centre-of-mass of Brown and York [2] and the ones based on Bramson's 
superpotential and the use of the holomorphic/anti-holomorphic 
spinors [3] at the spatial infinity. 

One of the most elegant introduction of the ADM conserved quantities 
at the spatial infinity is based on the requirement of the 
differentiability of the Hamiltonian. This approach of Regge and 
Teitelboim [4] was refined later by Beig and \'O Murchadha [5], 
recovering the ADM energy and linear momentum and the spatial angular 
momentum of Regge and Teitelboim, but giving a different expression 
for the centre-of-mass. The recent investigations of Baskaran, Lau 
and Petrov [6] show that the Brown--York centre-of-mass tends to the 
expression of Beig and \'O Murchadha. 

The traditional ADM approach of the conserved quantities and the 
Hamiltonian analysis of general relativity is based on the 3+1 
decomposition of the fields and the geometry. Hence it is not a 
priori clear that the energy and the spatial momentum form a Lorentz 
vector, or the spatial angular momentum and centre-of-mass form an 
anti-symmetric Lorentz tensor. To ensure the Lorentz covariance of 
the conserved quantities at spatial infinity Nester developed a 
spacetime-covariant Hamiltonian formulation of general relativity 
[7]. However, the {\it content} and the {\it results} of the theory 
can be spacetime covariant even if its form is not. Thus, in 
particular, we should find a {\it spacetime} interpretation of the 
traditional Hamiltonian formulation and the `conserved' quantities 
of the theory in terms of some appropriately defined asymptotic 
spacetime Killing vectors. 

It is known that the ADM energy-momentum can be finite and well 
defined (i.e. independent of the background structure) and the 
energy non-negativity can be proven even if the metric falls off 
with the radial distance $r$ slightly faster than $r^{-{1\over2}}$ 
[8-12]. This raises the question of finding the weakest possible 
fall-off for the metric and extrinsic curvature under which the 
spatial angular momentum and centre-of-mass are still finite and 
well defined. Furthermore, we should be able to treat not only the 
vacuum theory, but the matter fields should also be included. 

The present paper is devoted to the investigations of the global 
energy-momentum and angular momentum introduced at the spatial 
infinity of asymptotically flat spacetimes. We extend the analysis 
of canonical vacuum general relativity by R. Beig and N. \'O 
Murchadha [5] in numerous ways: the interpretatableness of the 
results in the {\it spacetime} is required, the symplectic structure 
is considered not to be fundamental and the emphasis is shifted to 
the field equations, the $1/r$ and $1/r^2$ {\it a priori} fall-off 
conditions for the metric and the canonical momentum, respectively, 
are relaxed, the matter fields are included and the background 
dependence of the angular momentum is investigated. In the literature 
several mathematically inequivalent model for the spatial infinity 
have been suggested (see e.g. [13-16]). However, the notion of 
asymptotic flatness at 
the spatial infinity based on a spacelike hypersurface is expected 
to be the weakest possible in the sense that in every reasonable 
model of spatial infinity the existence of such a hypersurface is 
expected. Thus, in the present paper, we do not use any specific 
model of infinity, and the notion of asymptotic flatness that we 
use is based on the existence of a certain spacelike hypersurface. 
Since there is some recent interest in higher dimensional 
(Lorentzian) models (see e.g. [17-18]), and since no extra effort 
is needed to do the analysis in general $m=n+1$ spacetime dimensions, 
we assume only that the dimension of the spacetime is $m\geq3$. 

In the traditional analysis the lapse and the shift are implicitly 
assumed to depend only on the spatial coordinates, but not on the 
time coordinate. However, the equations of motion allow their time 
dependence. It turns out that it is precisely this freedom that 
makes possible to give the {\it spacetime} interpretation of the 
Poincare algebra of the Hamiltonians found by Beig and \'O Murchadha. 
If we excluded the time dependence of the lapse and the shift then 
we would not be able to recover the boost Killing vectors even of 
the Minkowski spacetime. 

The philosophy of our analysis deviates slightly from the traditional 
one. It is the equations of motion that are considered to be 
fundamental, and the boundary conditions at infinity are required to 
be the slowest power-type fall-off conditions compatible with the 
evolution equations. Then the symplectic structure and the Hamiltonian 
are considered to be only as secondary structures. They are considered 
to be important only from the point of view of finding observables and, 
in particular, the conserved quantities, but not from the point of 
view of the boundary conditions. This departs from the philosophy of 
Regge and Teitelboim, where the boundary conditions and the 
Hamiltonian were determined in a single procedure from the regularity 
and differentiability of the Hamiltonian. Thus, having the boundary 
conditions been specified, the Hamiltonian $H$ must be chosen such 
that the correct field equations be recoverable as the flows 
corresponding to the Hamiltonian vector fields of $H$. Then the 
value of this Hamiltonian on the constraint surface will define 
the ADM quantities. 
Although the Beig--\'O Murchadha analysis was carried out for the 
{\it vacuum} Einstein theory, the value of their Hamiltonian on the 
constraint surface can be used to {\it define} the energy-momentum 
and angular momentum even in the presence of matter fields. We clarify 
in what sense these quantities are conserved, and how they depend on 
the flat background metric. 

First we determine the weakest possible power-type fall-off 
conditions for the energy-momentum tensor of the matter fields, the 
metric, the extrinsic curvature, the lapse and the shift which, 
together with the parity conditions of Regge and Teitelboim, are 
preserved by the energy-momentum conservation law $T^{ab}{}_{;b}=0$ 
and the evolution equations for the geometry.  In an $n+1$ 
dimensional spacetime they are of order $O(r^{-(n+1)})$, $O(r^{-k})$ 
for some $k>0$, $O(r^{-(k+1)})$, $O(r)$ and $O(r)$, respectively. The 
spacetime vector fields built from these allowed lapses and shifts 
will be called the {\it allowed time axes}. Then the asymptotic 
spacetime Killing vectors (with respect to an allowed time axis $\xi
^a$) are defined to be those vector fields for which the Killing 
operator is of order $O(r^{-k})$, and the space of these asymptotic 
Killing vectors will be denoted by ${\cal A}^K_\xi$. Its factor 
${\cal A}^K_\xi/{\cal G}^K_\xi$ by the subspace ${\cal G}^K_\xi
\subset{\cal A}^K_\xi$ (whose elements are the asymptotic Killing 
vectors with $O(r^{1-k})$ asymptotic behaviour) can be endowed with 
a Lie algebra structure in a natural way, and this is shown to be 
isomorphic to the Lorentz Lie algebra for slow fall-off $k\in(0,1)$, 
but for faster fall-off, $k\geq1$, it is the Poincare algebra. Thus 
the structure of the Lie algebra of the asymptotic symmetries is 
linked to the fall-off rate $k$ of the metric. 

One way of associating conserved quantities to asymptotically flat 
spacetimes is the use of the symplectic/Hamiltonian formalism. 
Since the details of the formalism depend on the type of the fields, 
we concentrate only on the vacuum theory. It is shown that the 
applicability of the symplectic formalism to the vacuum general 
relativity, in particular the existence of the symplectic 2-form, 
already implies that $k\geq{1\over2}(n-1)$. This excludes the slow 
($k<1$) fall-off in spacetime dimensions greater than 3. The 
constraint functions are shown to be finite, functionally 
differentiable and close to a Poisson algebra precisely for those 
lapses and shifts that correspond to the special allowed time axes 
with $O(r^{1-k})$ asymptotic behaviour. We show that the generators 
of the gauge transformations, i.e. the functions whose Hamiltonian 
vector fields span the kernel of the pull back to the constraint 
surface of the symplectic 2-form, are precisely these special time 
axes. A subspace ${\cal A}^0_\xi\subset{\cal A}^K_\xi$ is found such 
that the Hamiltonian $H$ of Beig and \'O Murchadha, mapping ${\cal 
A}^K_\xi$ into the Poisson algebra of functions, preserves the Lie 
product of the elements of ${\cal A}^0_\xi$. It is shown that 
$H[K^a]$ is constant in time with respect to the allowed time axes 
$\xi^a$ if $K^a\in{\cal A}^0_\xi$, and it is only constant modulo 
constraint functions for $K^a\in{\cal A}^K_\xi$. 

The boundary term in the Hamiltonian $H[K^a]$ is used to define the 
energy-momentum and angular momentum even in the presence of matter, 
independently of any symplectic structure. It is shown that, although 
the energy-momentum is well defined even for $k>{1\over2}(n-2)$, the 
angular momentum and centre-of-mass are finite only if $k\geq{1\over
2}(n-1)$. Similar result was obtained independently by Baskaran, Lau 
and Petrov in 3+1 dimensions recently [6]. $H[K^a]$ reproduces the 
ADM energy and spatial momentum, the spatial angular momentum of 
Regge and Teitelboim and the centre-of-mass of Beig and \'O Murchadha 
for $K^a\in{\cal A}^K_\xi$ if the allowed time axis $\xi^a$ 
corresponds to a gauge generator. 
However, the familiar boost Killing vectors of the Minkowski 
spacetime are contained in ${\cal A}^K_\xi$ only if the lapse part 
of $\xi^a$ does not tend to zero at infinity. For $K^a\in{\cal A}^K
_\xi$ with time axes $\xi^a$ describing pure time translation at 
infinity the Hamiltonian $H[K^a]$ reproduces the energy-momentum 
and spatial angular momentum above, but gives an additional term 
(the spatial momentum times the coordinate time) to the 
centre-of-mass of Beig and \'O Murchadha. To derive the familiar, 
expected transformation law for the (relativistic) angular momentum 
this extra term is needed. 
For $K^a\in{\cal A}^K_\xi$ the value of the Hamiltonian $H[K^a]$ 
is shown to be constant in time with respect to the allowed time axis 
$\xi^a$ provided the constraint equations are satisfied. 
We investigate the conditions of the background (in-)dependence of 
the energy-momentum and angular momentum, and we found that although 
the former is well defined even if the diffeomorphisms representing 
the ambiguity of the background metric tend to rigid Euclidean 
transformations as $O(r^R)$, where $R\leq-k$ and $R<(3-n)$, the 
latter is well defined if $R\leq(1-k)$ and $R\leq(2-n)$ (and in the 
case of the equality, $R=(2-n)$, the generator of the diffeomorphism 
has odd parity). In particular, to have well defined angular 
momentum in 3+1 dimensions the metric must fall off at least as 
$O(r^{-1})$, and the allowed diffeomorphisms must tend to rigid 
Euclidean transformations at least as $O(r^{-1})$. 

In subsection 2.1 the necessary tools are introduced and reviewed, 
mostly to fix the notations and conventions. The new key element 
here is the $n+1$ decomposition and analysis of the Killing operator. 
To motivate the boundary conditions and the precise definition of 
the asymptotic spacetime Killing vectors we discuss the Minkowski 
spacetime in subsection 2.2. Then, in subsections 2.3 and 2.4, the 
boundary conditions and the asymptotic Killing vectors are 
discussed. Section 3. is devoted to the analysis of the canonical 
general relativity, in particular to the constraints, the gauge 
transformations and the Hamiltonian. In section 4. we apply the 
Beig--\'O Murchadha Hamiltonian to define the energy-momentum and 
angular momentum of general asymptotically flat spacetimes, even 
in the presence of the matter fields, and clarify how these 
quantities depend on the background metric. The appendix is the 
brief discussion of the boundary conditions for the matter fields 
at the {\it null} infinity.

We use the abstract index formalism, and only the underlined and 
boldface indices take numerical values. The spacetime dimension and 
the signature will be assumed to be $m=n+1$ and $1-n$, respectively. 
The Riemann and Ricci tensors and the curvature scalar e.g. of the 
spacetime connection $\nabla_a$ are defined by $-{}^mR^a{}_{bcd}X^a
Y^cZ^d:=\nabla_Y(\nabla_ZX^a)-\nabla_Z(\nabla_YX^a)-\nabla_{[Y,Z]}X^a$, 
${}^mR_{ab}:={}^mR^c{}_{acb}$ and ${}^mR:={}^mR_{ab}g^{ab}$, 
respectively. Thus, Einstein's equations take the form ${}^mR_{ab}-{1
\over2}{}^mRg_{ab}+\Lambda g_{ab}=-\kappa T_{ab}$, and we use the 
units in which $c=1$. The terminology and formalism that we follow 
in the symplectic/Hamiltonian description of general relativity are 
mostly based on [19-22], and the level of the mathematical rigor 
we work at in section 3. corresponds to that of [21,22]. 

\bigskip
\bigskip

\ni
{\lbf 2. The $m=n+1$ decomposition}\par
\medskip
\ni
{\bf 2.1 The n+1 form of the Einstein equations and the Killing 
operator}\par
\medskip
\ni
Let $\Sigma_t$ be a smooth foliation of the $m=n+1$ dimensional 
spacetime $(M,g_{ab})$ by spacelike hypersurfaces and let $t^a$ be 
its future directed unit normal such that $t^a\nabla_at$ is positive. 
Let $P^a_b:=\delta^a_b-t^at_b$, the orthogonal projection to $\Sigma_t$. 
The lapse function $N$, the extrinsic curvature $\chi_{ab}$ and the 
acceleration $a_e$ of the foliation are defined by $Nt^a\nabla_at:=1$, 
$\chi_{ab}:=P^e_aP^f_b\nabla_et_f={1\over2}\L_{\bf t}q_{ab}$ and $a_e:=t^a
\nabla_at_e=-D_e(\ln N)$, respectively. Here $q_{ab}$ is the induced 
(negative definite) metric and $D_a$ is the corresponding intrinsic 
Levi-Civita covariant derivative. The corresponding Riemann and 
Ricci tensors and the curvature scalar will be denoted by $R^a{}
_{bcd}$, $R_{ab}$ and $R$, respectively. If $\xi^a$ is any smooth 
vector field such that $\xi^a\nabla_at=1$ (`evolution vector field' 
or rather `general time axis'), then $\xi^a$ has the form $\xi^a=N
t^a+N^a$ for some vector field $N^a=N^bP^a_b$, the shift part of $\xi
^a$. For any vector field $\xi^a$ and purely spatial tensor field 
$T^{a_1...a_r}_{b_1...b_s}$ let us define the `time derivative' of 
$T^{a_1...a_r}_{b_1...b_s}$ by 

$$
\dot T^{a_1...a_r}_{b_1...b_s}:=P^{a_1}_{e_1}...P^{a_r}_{e_r}P
^{f_1}_{b_1}...P^{f_s}_{b_s}\L_{\xi}T^{e_1...e_r}_{f_1...f_s}=
NP^{a_1}_{e_1}...P^{a_r}_{e_r}P^{f_1}_{b_1}...P^{f_s}_{b_s}\L
_{\bf t}T^{e_1...e_r}_{f_1...f_s}+\L_{\bf N}T^{a_1...a_r}_{b_1...
b_s}.\eqno(2.1.1)
$$
\ni
In particular, for the time derivative of the induced metric we 
have 

$$
\dot q_{ab}=2N\chi_{ab}+\L_{\bf N}q_{ab}.\eqno(2.1.2)
$$
\ni
Let $\varepsilon_{a_1...a_m}$ be the spacetime volume $m$-form. The 
induced volume $n$-form and volume element on $\Sigma_t$ is defined 
by $\varepsilon_{a_1...a_n}:=t^a\varepsilon_{aa_1...a_n}$ and ${\rm d}
\Sigma_t:={1\over n!} \varepsilon_{a_1...a_n}=\sqrt{\vert q\vert}{\rm 
d}^nx$, respectively, and hence ${\rm d}v=N{\rm d}\Sigma_t{\rm d}t$. 
(The other convention for the orientation of the submanifolds, which 
would be slightly more convenient if unitary spinors were used 
[23], is when $\varepsilon_{a_1...a_n}:=\varepsilon_{a_1...a_na}t^a=
(-)^nt^a\varepsilon_{aa_1...a_n}$.) If $B\subset\Sigma$ is a compact 
$n$ dimensional submanifold with smooth boundary ${\cal S}$, $v^a$ 
its outward directed unit normal in $\Sigma$, then, by the negative 
definiteness of $q_{ab}$, for any vector field $X^a$ tangent to 
$\Sigma$ the Gauss law takes the form $\int_BD_bX^b{\rm d}\Sigma=-
\oint_{\cal S}X^bv_b{\rm d}{\cal S}$, where ${\rm d}{\cal S}:={1\over
(n-1)!}t^ev^f\varepsilon_{efa_2...a_n}$. 

The conservation of the matter energy-momentum, i.e. $T^{ab}{}_{;b}=
0$, is equivalent to 

$$\eqalignno{
\dot\mu&=N\Bigl(-D_ej^e+\sigma^{ab}\chi_{ab}-{2\over N}j^eD_eN-\mu\chi
  \Bigr)+\L_{\bf N}\mu, &(2.1.3)\cr
\dot j^b&=N\Bigl(-D_a\sigma^{ba}-{1\over N}\sigma^{ba}D_aN+\mu{1\over 
  N}D^bN-2j_a\chi^{ab}-\chi j^b\Bigr)+\L_{\bf N}j^b,  &(2.1.4)\cr}
$$
\ni
where $\mu:=T^{ab}t_at_b$, $j^b:=T^{ae}t_aP^b_e$ and $\sigma^{ab}:=T
^{ef}P^a_eP^b_f$. The projections of the $m$ dimensional Einstein 
equations, ${}^mG_{ab}+\Lambda g_{ab}+\kappa T_{ab}=0$, are the 
constraints 

$$\eqalignno{
&\kappa c:=t^at^b\Bigl({}^mG_{ab}+\Lambda g_{ab}+\kappa T_{ab}\Bigr)
 =-{1\over2}\Bigl(R+\bigl(\chi^2-\chi_{ab}\chi^{ab}\bigr)\Bigr)
 +\Lambda+\kappa\mu=0, &(2.1.5)\cr
&\kappa c_a:=P^e_at^b\Bigl({}^mG_{eb}+\Lambda g_{eb}+\kappa T_{eb}
 \Bigr)=-\Bigl(D_e\chi^e{}_a-D_a\chi\Bigr)+\kappa j_a=0; &(2.1.6)\cr}
$$
\ni
and the evolution equations 

$$\eqalign{
\dot\chi_{cd}&=N\Bigl(-R_{cd}+2\chi_{ce}\chi^e{}_d-\chi\chi_{cd}\Bigr)
 +\L_{\bf N}\chi_{cd}-D_cD_dN+\cr
&+{2\over(n-1)}\Lambda Nq_{cd}+\kappa N\Bigl(-\sigma_{cd}+{1\over(n
 -1)}\sigma^e{}_eq_{cd}+{1\over(n-1)}\mu q_{cd}\Bigr).\cr}\eqno(2.1.7)
$$
\ni
To check whether the evolution equations (2.1.2) and (2.2.7) 
preserve the constraints, take the time derivative of $c$ and $c_a$ 
and use (2.1.2)--(2.1.7). We get 

$$\eqalignno{
\dot c&=-2c^aD_aN-ND_ac^a+\L_{\bf N}c-2N\chi c, &(2.1.8)\cr
\dot c_a&=2cD_aN+ND_ac+\L_{\bf N}c_a-N\chi c_a. &(2.1.9)\cr}
$$
\ni
Therefore, if the constraints (2.1.5), (2.1.6) are satisfied at 
$t=0$, then any of their derivatives also vanish, and hence the 
constraints are preserved by the evolution equations (2.1.2), 
(2.1.7). Since in the present paper we are interested in 
asymptotically flat spacetimes, the cosmological constant $\Lambda$ 
will be assumed to be zero. 

If $K^a=:Mt^a+M^a$ is any smooth vector field, where $M^a=P^a_bM^b$, 
then the $n+1$ decomposition of the so-called Killing operator, 
acting on the spacetime 1-form field $K_a$, is 

$$\eqalignno{
Nt^at^b\nabla_{(a}K_{b)}&=\dot M+M^aD_aN-N^aD_aM,&(2.1.10)\cr
2NP^b_at^c\nabla_{(b}K_{c)}&=\dot M_a+\Bigl(ND_aM-MD_aN\Bigr)
  -2N\chi_{ab}M^b-\L_{\bf N}M_a,&(2.1.11)\cr
P^c_aP^d_b\nabla_{(c}K_{d)}&=D_{(a}M_{b)}+M\chi_{ab}.&(2.1.12)\cr}
$$
\ni
Clearly, while the space--space projection of the Killing operator 
is well defined even on a single hypersurface $\Sigma$, the first 
two projections are well defined only if a foliation $\Sigma_t$ of 
$M$, i.e. a lapse function $N$ on $\Sigma$, is fixed, and, in addition, 
their right hand side needs a choice for the shift vector $N^a$ too. 
Obviously, in a generic spacetime the Killing equation $\nabla_{(a}
K_{b)}=0$ has only the trivial solution. However, $t^at^b\nabla_{(a}
K_{b)}=0$ and $P^b_at^c\nabla_{(b}K_{c)}=0$ can always be solved, 
i.e. the initial value problem for the system 

$$\eqalignno{
\dot M=&-M^aD_aN+N^aD_aM,&(2.1.13)\cr
\dot M_a=&-\Bigl(ND_aM-MD_aN\Bigr)+2N\chi_{ab}M^b+\L_{\bf N}M_a,
&(2.1.14)\cr}
$$
\ni
is unconstrained, and the initial value problem for $M$ and $M^a$ 
always has a solution. If $\bar K^a=\bar Mt^a+\bar M^a$ is another 
spacetime vector field, then the $n+1$ decomposition of the 
Lie-bracket of $K^a$ and $\bar K^a$ is 

$$\eqalign{
\Bigl[K,\bar K\Bigr]^a=&\bigl(t^at^b+2q^{ab}\bigr)\Bigl(M\nabla_{(b}
 \bar K_{c)}-\bar M\nabla_{(b}K_{c)}\Bigr)t^c+\cr
+&t^a\Bigl(\L_{\bf M}\bar M-\L_{\bar{\bf M}}M\Bigr)+\Bigl[M,\bar M
  \Bigr]^a+\Bigl(\bar MD^aM-MD^a\bar M\Bigr).\cr}\eqno(2.1.15)
$$
\ni
Thus in this decomposition only the time-time and the time-space 
parts of the Killing operator appear, but not the space-space 
parts. Therefore, if both $K^a$ and $\bar K^a$ satisfied (2.1.13-14), 
then the projections of the Killing operator on the right hand 
side of (2.1.15) would be vanishing. 

\bigskip
\bigskip

\ni
{\bf 2.2 Boundary conditions I.: Matter fields in Minkowski spacetime}
\par
\medskip
\ni
Let $(M,g_{ab})$ be the Minkowski spacetime and ${\cal K}$ the Lie 
algebra of its Killing vectors. As is well known, it contains an $m$ 
dimensional commutative ideal ${\cal I}$, consisting of the constant 
vector fields on $M$ and inheriting a natural Lorentzian vector 
space structure too, and ${\cal K}/{\cal I}\approx so(1,n)$. Fixing a 
Cartesian coordinate system $X^{\ua}=(\tau,X^{\bi})$, ${\ua}=0,1,...,n$ 
and ${\bi}=1,2,...,n$, (i.e. adapting the coordinates $X^{\ua}$ to an 
orthonormal basis of the space of the constant vector fields), the 
translation and boost-rotation Killing 1-forms are well known to 
take the form $K^{\ua}_e=\nabla_eX^{\ua}$ and $K^{{\ua}{\ub}}_e=X^{\ua}
\nabla_eX^{\ub}-X^{\ub}\nabla_eX^{\ua}$, respectively. Therefore, any 
Killing 1-form can be written in the form $K_e=R_{\ua\ub}K^{\ua\ub}_e+
T_{\ua}K^{\ua}_e=R_{\bi\bj}(X^{\bi}\nabla_eX^{\bj}-X^{\bj}\nabla_eX^{\bi})
+2B_{\bi}(X^{\bi}\nabla_e\tau-\tau\nabla_eX^{\bi})+T_{\bi}\nabla_eX^{\bi}
+T\nabla_e\tau$, where $R_{\ua\ub}=-R_{\ub\ua}$ and $T_{\ua}$ are 
constant, and we used the notations $B_{\bi}:=R_{\bi 0}$ and $T:=T_0$.

Let $\Sigma_\tau$ be a $\tau={\rm const}$ hyperplane, $R^2:=\delta_{\bi
\bj}X^{\bi}X^{\bj}$, and for some $R>0$ let $B_R$ be the solid closed 
ball of $\Sigma_\tau$ with radius $R$ and ${\cal S}_R:=\partial B_R$ 
its boundary. If $v^a$ is its outward directed unit normal, then $1=v^a
D_aR=-{X^{\bi}\over R}q_{\bi\bj}v^aD_aX^{\bj}$, where $q_{\bi\bj}$ are 
the components of the induced flat metric $q_{ab}$ on $\Sigma_\tau$ in 
the coordinate system $\{X^{\bi}\}$, and hence $v^{\bi}={X^{\bi}\over 
R}$. If $f=f(\tau,R,{X^{\bi}\over R})$ is any function, 
then let us define its even and odd parity parts, respectively, by ${}
^\pm f(\tau,R,{X^{\bi}\over R}):={1\over2}(f(\tau,R,{X^{\bi}\over R})
\pm f(\tau,R,-{X^{\bi}\over R}))$. Let $\tau_a$ be the future pointing 
unit timelike normal to $\Sigma_\tau$, and define the quasi-local 
energy, spatial momentum, spatial angular momentum and centre-of-mass 
of the matter fields in the n-ball $B_R$, respectively, by taking the 
flux integral of the conserved current $K_aT^{ab}$: 

$$\eqalignno{
{\tt E}_R:=&\int_{B_R}K^0_aT^{ab}\tau_b{\rm d}\Sigma=\int_0^R\Bigl(
 \oint_{\cal S}\mu\,{\rm d}{\cal S}\Bigr){R'}^{n-1}{\rm d}R'=\int_0^R
 \Bigl(\oint_{\cal S}{}^+\mu\,{\rm d}{\cal S}\Bigr){R'}^{n-1}{\rm d}R', 
 &(2.2.1.a)\cr
{\tt P}^{\bi}_R:=&\int_{B_R}K^{\bi}_aT^{ab}\tau_b{\rm d}\Sigma=\int_0
 ^R\Bigl(\oint_{\cal S}j^aD_aX^{\bi}{\rm d}{\cal S}\Bigr){R'}^{n-1}
 {\rm d}R'=\int_0^R\Bigl(\oint_{\cal S}{}^+j^{\bi}{\rm d}{\cal S}
 \Bigr){R'}^{n-1}{\rm d}R', &(2.2.1.b)\cr
{\tt J}^{\bi\bj}_R:=&\int_{B_R}K^{\bi\bj}_aT^{ab}\tau_b{\rm d}\Sigma=
 \int_0^R\Bigl(\oint_{\cal S}2j^av^{[{\bi}}D_aX^{{\bj}]}{\rm d}{\cal 
 S}\Bigr){R'}^n{\rm d}R'=-2\int_0^R\Bigl(\oint_{\cal S}{}^-j^{[{\bi}}
 v^{{\bj}]}\,{\rm d}{\cal S}\Bigr){R'}^n{\rm d}R', &(2.2.1.c)\cr
{\tt J}^{\bi 0}_R:=&\int_{B_R}K^{\bi 0}_aT^{ab}\tau_b{\rm d}\Sigma=
 \int_0^R\Bigl(\oint_{\cal S}\mu v^{\bi}{\rm d}{\cal S}\Bigr){R'}^n
 {\rm d} R'-\tau{\tt P}^{\bi}=\int_0^R\Bigl(\oint_{\cal S}{}^-\mu v
 ^{\bi}{\rm d}{\cal S}\Bigr){R'}^n{\rm d}R'-\tau{\tt P}^{\bi}, 
 &(2.2.1.d)\cr}
$$
\ni
where ${\rm d}{\cal S}$ is the area element on the unit sphere ${\cal 
S}$. (Strictly speaking, the traditional [non-conserved 
non-relativistic] centre-of-mass is ${\tt J}^{\bi 0}+\tau\,{\tt P}
^{\bi}$.) Since $K_aT^{ab}$ is divergence-free for Killing vectors, 
these quasi-local quantities are, in fact, associated with the 
$(n-1)$-surface ${\cal S}_R$ and depend only on $K_a$: if $\tilde
\Sigma$ is any compact spacelike hypersurface whose boundary $\partial
\tilde\Sigma$ coincides with ${\cal S}_R$, then the flux integral of 
$T^{ab}K_b$ on $\tilde\Sigma$ will be that on $B_R$. The necessary 
and sufficient condition of the existence of the $R\rightarrow\infty$ 
limit of these integrals, respectively, 
is\footnote{${}^1$}{A function $f(r)$ will be called of order $o(r
^{-k})$ if $\lim_{r\rightarrow\infty}(f(r)r^k)=0$, and will be called 
of order $O(r^{-k})$ if $\lim_{r\rightarrow\infty}(f(r)r^k)$ exists. 
In particular, $o(r^{+0})$ will denote logarithmic divergence and $o(r
^{-0})$ logarithmic fall-off, while $O(1)$ means that $f(r)$ tends to 
a constant at infinity.} 

$$\eqalignno{
&R\oint_{{\cal S}_R}\,\mu{\rm d}{\cal S}_R=R^n\oint_{\cal S}\,{}^+
 \mu{\rm d}{\cal S}=o(R^{-0}),&(2.2.2.a)\cr
&R\oint_{{\cal S}_R}\,j^aD_aX^{\bi}{\rm d}{\cal S}_R=R^n\oint_{\cal S}
 \,{}^+j^{\bi}{\rm d}{\cal S}=o(R^{-0}),&(2.2.2.b)\cr
&2R^2\oint_{{\cal S}_R}\,j^av^{[{\bi}}D_aX^{{\bj}]}{\rm d}{\cal S}_R=
 -2R^{(n+1)}\oint_{\cal S}{}^-j^{[{\bi}}v^{{\bj}]}{\rm 
 d}{\cal S}=o(R^{-0}),&(2.2.2.c)\cr
&2R^2\oint_{{\cal S}_R}\,\mu v^{\bi}{\rm d}{\cal S}_R=2R^{(n+1)}\oint
 _{\cal S}\,{}^-\mu v^{\bi}{\rm d}{\cal S}=o(R^{-0}).
 &(2.2.2.d)\cr}
$$
\ni
These {\it global integral conditions} can be ensured by the {\it 
explicit} fall-off and parity conditions 

$$\eqalignno{
\mu(\tau,R,{X^{\bk}\over R})&={1\over R^m}\,{}^+\mu^{(m)}(\tau,{X^{\bk}
 \over R})+o(R^{-m}),&(2.2.3.a)\cr
j^{\bi}(\tau,R,{X^{\bk}\over R})&={X^{\bi}\over R}\,{}^+f(\tau,R,{X
 ^{\bk}\over R})+{1\over R^m}\,{}^+j^{{\bi}(m)}(\tau,{X^{\bk}\over R})
 +o(R^{-m}),\cr}
$$
\ni
where ${}^+f(\tau,R,{X ^{\bk}\over R})$ is an arbitrary function with 
even parity. $\mu^{(m)}$ and $j^{{\bi}(m)}$ contribute only to the 
energy and the spatial momentum but not to the angular momentum and 
centre-of-mass, hence we may call them the {\it ADM mass aspect} of 
$T^{ab}$. Repeating this analysis on boosted hyperplanes $\Sigma'
_{\tau'}:=\{\tau':=\tau\cosh\beta+X^{\bi}\alpha_{\bi}\sinh\beta={\rm 
const}\}$, where $\beta\in{\bf R}$ and $q^{\bi\bj}\alpha_{\bi}\alpha
_{\bj}=-1$, we obtain that ${}^+f(\tau,R,{X^{\bk}\over R})=0$ and, 
in addition to (2.2.3a), 

$$\eqalignno{
j^{\bi}(\tau,R,{X^{\bk}\over R})&={1\over R^m}\,{}^+j^{{\bi}(m)}(\tau,
 {X^{\bk}\over R})+o(R^{-m}), &(2.2.3.b)\cr
\sigma^{\bi\bj}(\tau,R,{X^{\bk}\over R})&={1\over R^m}\,{}^+\sigma^{\bi
 \bj}{}^{(m)}(\tau,{X^{\bk}\over R})+o(R^{-m}).&(2.2.3.c)\cr}
$$
\ni
Note that although (2.2.2) could be ensured only by fall-off 
conditions that are strictly faster than those in (2.2.3), but with 
these conditions we would exclude e.g. the electromagnetic field from 
our investigations, where the typical fall-off of the energy-momentum 
tensor is $R^{-m}$. Obviously, the fall-off and parity conditions 
(2.2.3) are only {\it sufficient}, and the global integral conditions 
(2.2.2) can be satisfied without the parity conditions too. The 
advantage of the fall-off and parity conditions is that they can be 
given {\it explicitly}. However, the price that we had to pay for this 
is that we excluded those field configurations from our investigations 
that satisfy the global integral conditions (2.2.2) but not the 
explicit fall-off and parity conditions (2.2.3). 

One can carry out a similar analysis of the fall-off and global 
integral conditions that can ensure the finiteness of the global 
energy-momentum and (relativistic) angular momentum at the future 
{\it null infinity}. Since, however, in the present paper primarily 
we are interested in the kinematical quantities defined at the {\it 
spatial} infinity, the null infinity case will be discussed only in 
the Appendix.

In the rest of this subsection we discuss the conditions under which 
the ${1\over2}m(m+1)$ spacetime Killing vectors can be recovered, at 
least asymptotically, from quantities defined on a general 
asymptotically flat spacelike hypersurface in the Minkowski spacetime. 
Since, however, many parts of the following discussion are well known 
from various sources, we sketch only the main points of the 
argumentation. 

The global Cartesian coordinate system $X^{\ua}=(\tau,X^{\bi})$ defines 
a foliation of the Minkowski spacetime by the hyperplanes $\Sigma_\tau$ 
and gives the `time axis' $({\partial\over\partial\tau})^a$, i.e. the 
corresponding lapse is one and the shift is zero. Thus our aim is to 
determine those conditions under which a coordinate system $(t,x
^{\bi})$, based on a more general spacelike hypersurface $\Sigma$, 
`approaches asymptotically' the Cartesian coordinate system `at 
infinity'. However, to do `a coordinate system approaches a Cartesian 
coordinate system asymptotically at infinity' to be meaningful we 
need to use some model of the spacelike infinity of the Minkowski 
spacetime. We choose the classical conformal boundary of Penrose. 
Thus let $(\tilde M,\tilde g_{ab})$ be the conformally compactified 
Minkowski spacetime (e.g. the closure of the conformally embedded 
Minkowski spacetime in the Einstein universe, see e.g. [24]), 
$\Omega$ the conformal factor on $\tilde M$ such that $\tilde g_{ab}
\vert_M=\Omega^2\vert_M\,g_{ab}$ and ${\rm i}^0\in\tilde M$ the point 
on the conformal boundary of $(M,g_{ab})$ representing the spatial 
infinity. Then the family of hyperplanes $\Sigma_\tau$ uniquely 
determines a family $\tilde\Sigma_\tau$ of smooth Cauchy surfaces in 
$\tilde M$ such that ${\rm i}^0\in\tilde\Sigma_\tau$ for all $\tau\in
{\bf R}$, and the future directed $\tilde g_{ab}$-unit normal $\tilde
\tau_a$ of all the surfaces $\tilde\Sigma_\tau$ coincide at ${\rm i}
^0$. The compactification of $\Sigma_\tau$ to $\tilde\Sigma_\tau$ can 
also be done by the standard inversion transformation: on some open 
neighbourhood $\tilde V_\tau$ of ${\rm i}^0$ in $\tilde\Sigma_\tau$ 
let $\tilde X^{\bi}:=R^{-2}X^{\bi}$. Then the coordinates $\tilde X
^{\bi}$ can be extended from $\tilde V_\tau-\{{\rm i}^0\}$ to $\tilde 
V_\tau$ by $\tilde X^{\bi}({\rm i}^0)=0$, and $\tilde E^a_{\bi}:=(
{\partial\over\partial\tilde X^{\bi}})^a\vert_{{\rm i}^0}$ is a 
$\tilde g_{ab}$-orthonormal spatial basis at ${\rm i}^0$ orthogonal 
to $\tilde\tau_a$. This basis turns out to be independent of $\tau$. 

Let $\tilde\Sigma$ be any smooth spacelike hypersurface in $\tilde 
M$ through ${\rm i}^0$. We say that $\Sigma:=\tilde\Sigma-\{{\rm i}^0
\}$ is approaching the leaves of the foliation $\Sigma_\tau$ at 
infinity if the future directed $\tilde g_{ab}$-unit normal $\tilde t
_a$ of $\tilde\Sigma$ coincides with $\tilde\tau_a$ at ${\rm i}^0$. 
We assume that $\Sigma$ is such a hypersurface, otherwise it is called 
asymptotically boosted with respect to the leaves $\Sigma_\tau$. (Note 
that we have to assume the smoothness of $\tilde\Sigma$ even at ${\rm 
i}^0$, because otherwise its normal would not be well defined.) Let 
$t_a$ be the future directed $g_{ab}$-unit normal to $\Sigma$, and let 
$q_{ab}$ and $\tilde q_{ab}$ be the induced metrics and $\chi_{ab}$ 
and $\tilde\chi_{ab}$ the extrinsic curvatures of $\Sigma$ and $\tilde
\Sigma$, respectively. Then, by the construction, $(\tilde\Sigma,{\rm 
i}^0,\Omega\vert_{\tilde\Sigma},\tilde q_{ab},\tilde\chi_{ab})$ is an 
asymptote for $(\Sigma,q_{ab},\chi_{ab})$ in the sense of [13] (see 
also [25]), and, in addition, the normal directional derivative of 
the conformal factor, $\dot\Omega:=t^a\nabla_a\Omega\vert_\Sigma$, has 
a $C^3$ extension to $\tilde\Sigma$ such that $\dot\Omega({\rm i}^0)=0$, 
$(D_a\dot\Omega)({\rm i}^0)=0$, $(D_aD_b\dot\Omega)({\rm i}^0)=0$ and 
$(D_aD_bD_c\dot\Omega)({\rm i}^0)=0$, and $\tilde\chi_{ab}\vert_{\Sigma}
=\Omega\chi_{ab}+\dot\Omega q_{ab}$ holds. 
Let $\{\tilde x^{\bk}\}$ be a local coordinate system on some open 
neighbourhood $\tilde U$ of ${\rm i}^0$ in $\tilde\Sigma$ such that 
$\tilde x^{\bk}({\rm i}^0)=0$, and in these coordinates $({\partial
\over\partial\tilde x^{\bi}})^a\vert_{{\rm i}^0}=\tilde E^a_{\bi}$ and 
$\tilde\Gamma^{\bi}_{\bj\bk}({\rm i}^0)=0$ hold (i.e., in particular, 
$\tilde q_{\bi\bj}({\rm i}^0)=-\delta_{\bi\bj}$ holds, and $\{\tilde x
^{\bk}\}$ is a normal coordinate system with origin ${\rm i}^0\in
\tilde\Sigma$). Let $\tilde r^2:=\delta_{\bi\bj}\tilde x^{\bi}\tilde x
^{\bj}$, and define the new coordinates $x^{\bi}:=\tilde r^{-2}\tilde 
x^{\bi}$ and radial coordinate distance $r^2:=\delta_{\bi\bj}x^{\bi}
x^{\bj}=\tilde r^{-2}$ on $U:=\tilde U-\{{\rm i}^0\}$. (In general the 
coordinates $x^{\bi}$ are {\it not} the restrictions to $\Sigma$ of 
the Cartesian coordinates $X^{\bi}$.) The properties of $\Omega\vert
_{\tilde\Sigma}$ and $\dot\Omega$ stated above imply that in these 
coordinates $\Omega=r^{-2}(1+O(r^{-k}))$ and $\dot\Omega=O(r^{-(3+
h)})$ for some positive $k$ and $h$. (In fact, even for general 
asymptotically flat spacetimes when $\tilde\Sigma$ is not a smooth 
hypersurface at ${\rm i}^0$, one has $k,h\geq1$.) Then for the metrics 
and the extrinsic curvatures we have 

$$\eqalign{
q_{\bi\bj}{\rm d}x^{\bi}{\rm d}x^{\bj}=&\Omega^{-2}\tilde q_{\bi\bj}
 {\rm d}\tilde x^{\bi}{\rm d}\tilde x^{\bj}=\Bigl(-\delta_{\bi\bj}+{1
 \over r^k}q^{(k)}_{\bi\bj}+O\bigl(r^{-2}\bigr)+o\bigl(r^{-k}\bigr)
 \Bigr){\rm d}x^{\bi}{\rm d}x^{\bj}, \cr
\chi_{\bi\bj}{\rm d}x^{\bi}{\rm d}x^{\bj}=&\Bigl(\Omega^{-1}\tilde
 \chi_{\bi\bj}-\Omega^{-3}\dot\Omega\tilde q_{\bi\bj}\Bigr){\rm d}
 \tilde x^{\bi}{\rm d}\tilde x^{\bj}=\Bigl({1\over r^{1+h}}\chi^{(1+
 h)}_{\bi\bj}+o\bigl(r^{-(1+h)}\bigr)\Bigr){\rm d}x^{\bi}{\rm d}x
 ^{\bj}\cr}
$$
\ni
for some $q^{(k)}_{\bi\bj}$ and $\chi^{(1+h)}_{\bi\bj}$ depending only 
on ${x^{\bk}\over r}$. Actually, for the Minkowski spacetime, both $k$ 
and $h$ must be greater than or equal to 2. Therefore, $\{x^{\bi}\}$ 
is an `asymptotically Cartesian' coordinate system with respect to 
$q_{ab}$, and, on $U$, it defines a (negative definite) flat metric 
${}_0q_{ab}$ with respect to which $\{x^{\bi}\}$ is Cartesian. 

To complete $\{x^{\bi}\}$ to a spacetime coordinate system $\{t,x
^{\bi}\}$ (at least on an open neighbourhood of $U\subset\Sigma$ in 
$M$), we need a whole family $\Sigma_t$ of such hypersurfaces 
providing a foliation of this neighbourhood. However, to ensure 
that this spacetime coordinate system (and not only $\{x^{\bi}\}$ 
on the single hypersurface $\Sigma_0=\Sigma$) approaches the 
Cartesian one, all the leaves $\Sigma_t$ of the foliation must 
approach the leaves $\Sigma_\tau$ at infinity, i.e. the future 
directed $\tilde g_{ab}$-unit normal $\tilde t_a(t)$ of $\tilde
\Sigma_t$ must coincide with $\tilde\tau_a$ at ${\rm i}^0$ for all 
$t\in{\bf R}$. (Otherwise the foliation $\Sigma_t$ would be 
asymptotically accelerating at ${\rm i}^0$ rather than being 
inertial.) Then there is a natural extension of the spatial 
coordinates from $\Sigma$ to all the leaves $\Sigma_t$ via the 
construction above: 
on $\Sigma_t$ let $\{x^{\bi}\}$ be the inversion of the normal 
coordinates $\tilde x^{\bi}$ on $\tilde\Sigma_t$ based on the basis 
$\{\tilde E^a_{\bi}\}$ at ${\rm i}^0$ and with the origin at ${\rm i}
^0$. Denoting the tangent of the curves $\gamma(t):=(t,x^{\bi})$, 
$x^{\bi}={\rm const.}$, by $\xi^a$ and defining the lapses and shifts 
by $\xi^a=Nt^a+N^a=\tilde N\tilde t^a+\tilde N^a$, we get that $N=
\Omega^{-1}\tilde N=O(r^{2-p})$ and $N^a=\tilde N^a=O(r^{-p})$, where 
$p\geq2$, because the leaves $\tilde\Sigma_t$ of the foliation are 
tangent to each other at ${\rm i}^0$. Therefore, we can write $N=N
^{(0)}({x^{\bk}\over r})+O(r^{-1})$. 

Since the leaves both of the foliations $\tilde\Sigma_t$ and $\tilde
\Sigma_\tau$ are tangent to each other at ${\rm i}^0$ and both $\tilde 
X^{\bi}$ and $\tilde x^{\bi}$ are normal coordinates based on the same 
basis $\tilde E^a_{\bi}$ at ${\rm i}^0$, $\tilde X^{\bi}=\tilde x
^{\bi}+O(\tilde r^p)$, 
where $p\geq2$. But since $R^2=r^2(1+O(r^{-1}))$ we have $X^{\bi}=x
^{\bi}+O(r^{2-p})$, implying that $({\partial\over\partial x^{\bi}})^a
=({\partial\over\partial X^{\bi}})^a+O(r^{1-p})$. Furthermore, the 
angle between the normals $\tilde\tau_a$ and $\tilde t_a$ of $\tilde
\Sigma_\tau$ and $\tilde\Sigma_t$ is of order $\tilde r^{p-1}$ if the 
leaves of the foliations are tangent to each other in the $(p-1)$th 
order. Thus the (hyperbolic) cosine of this angle is $t_a\tau^a=
\tilde t^a\tilde\tau^b\tilde g_{ab}=1+O(\tilde r^{p-1})$, implying that 
$({\partial\over\partial t})^a=\xi^a=Nt^a+N^a=N({\partial\over\partial
\tau})^a+O(r^{3-2p})+O(r^{-p})$. Therefore, the coordinate basis 
vectors $({\partial\over\partial x^{\bi}})^a$ tend to $({\partial\over
\partial X^{\bi}})^a$ asymptotically, but $({\partial\over\partial t})
^a$ tends to $({\partial\over\partial\tau})^a$ only if $p=2$, and the 
lapse has the form $N=1+O(r^{-1})$ (whenever $\tau=t+c+O(r^{-1})$, 
where $c$ is a constant). Expanding the 1-form basis $(\nabla_e\tau,
\nabla_eX^{\bi})$ in terms of $(t_e,D_ex^{\bi})$, where $D_e$ is the 
induced derivative operator in the leaves $\Sigma_t$, we can rewrite 
the general Killing vector of the Minkowski spacetime given in the 
first paragraph of this subsection. We obtain $K_e=R_{\bi\bj}(x^{\bi}
D_ex^{\bj}-x^{\bj}D_ex^{\bi})+2B_{\bi}(x^{\bi}t_e-tD_ex^{\bi})+s_{\bi}
D_ex^{\bi}+st_e$, where $s(t,x^{\bk})=s^{(0)}(t,{x^{\bk}\over r})+O(r
^{-1})$ and $s_{\bi}(t,x^{\bk})=s^{(0)}_{\bi}(t,{x^{\bk}\over r})+
O(r^{-1})$. Its structure is similar to that of $K_e$ in the Cartesian 
coordinates, but instead of the constant components of the 
translations, $s$ and $s_{\bi}$ are functions of $t$, ${x^{\bk}\over 
r}$ and higher powers of $r^{-1}$. Thus they analogous to the 
supertranslations in the BMS group of the null infinity, and hence 
it is natural to call $st^a$ and $s^{\bi}({\partial\over\partial x
^{\bi}})^a$ temporal and spatial supertranslations, respectively. 

To summarize: although the global energy-momentum of the matter 
fields in Minkowski spacetime can be ensured to be finite by the 
$R^{-m}$ fall-off conditions, both at the spatial and the null 
infinity, to have finite angular momentum and centre-of-mass {\it 
additional global integral conditions must also be imposed on the 
mass aspect of $T^{ab}$}. At the spatial infinity these global 
integral conditions can be ensured by explicit parity conditions. 
To be able to recover the familiar Killing vectors in their usual 
form on a general asymptotically flat spacelike hypersurface 
$\Sigma$, at least asymptotically, the lapse $N$, defining the time 
coordinate $t$, must tend to 1 as $r\rightarrow\infty$.

\bigskip
\bigskip
\ni
{\bf 2.3 Boundary conditions II.: Asymptotically flat spacetimes}\par
\bigskip
\ni
Suppose that $\Sigma$ 
is asymptotically Euclidean in the sense that for some compact 
subset $K\subset\Sigma$ the complement $\Sigma-K$ is diffeomorphic to 
a finite disjoint union of manifolds $\Sigma_{(i)}$, each of which 
is diffeomorphic to ${\bf R}^n-B$, where $B$ is a solid ball in ${\bf 
R}^n$. The pieces $\Sigma_{(i)}$ are called the asymptotic ends of 
$\Sigma$. Since the next analysis can be repeated on each $\Sigma
_{(i)}$, for the sake of simplicity we assume that there is only one 
such end. Suppose that there is a (negative definite) metric ${}_0q
_{ab}$ on $\Sigma$ such that it is flat on the asymptotic end $\Sigma
-K$. Let $\{x^{\bi}\}$ be a coordinate system on $\Sigma-K$ which is 
Cartesian with respect to ${}_0q_{ab}$, $r^2:=\delta_{\bi\bj}x^{\bi}x
^{\bj}$, the radial distance function with respect to ${}_0q_{ab}$, 
and let ${}_0D_e$ be the Levi-Civita covariant derivative operator 
corresponding to ${}_0q_{ab}$. Then the quotients ${x^{\bi}\over r}$ 
can be interpreted as coordinates both on the unit sphere ${\cal S}
\approx S^{n-1}$ and the sphere ${\cal S}_r$ of large coordinate 
radius $r$ in $\Sigma-K$, and the components ${}_0v^{\bi}$ of the 
outward directed ${}_0q_{ab}$-unit normal ${}_0v^a$ to ${\cal S}_r$ 
in the coordinate system $\{x^{\bk}\}$ too. By a ball of radius $r$ 
in $\Sigma$ we mean $B_r:=\{\,p\in\Sigma-K\,\vert\,r(p)\leq r\,\}\cup 
K$. 

Let us consider the first, intuitively obvious condition of 
asymptotic flatness on the components of the metric and extrinsic 
curvature in the coordinate system $\{x^{\bi}\}$: for some positive 
$k$ and $l$ 

$$\eqalignno{
q_{\bi\bj}(x^{\bk})&={}_0q_{\bi\bj}+{1\over r^k}q_{\bi\bj}{}^{(k)}
 ({x^{\bk}\over r})+o(r^{-k}),&(2.3.1a)\cr
\chi_{\bi\bj}(x^{\bk})&={1\over r^l}\chi_{\bi\bj}{}^{(l)}({x^{\bk}
 \over r})+o(r^{-l}).&(2.3.1b)\cr}
$$
\ni
Therefore, the coefficients $q_{\bi\bj}{}^{(k)}$, $\chi_{\bi\bj}{}
^{(l)}$ can be interpreted as functions defined only on the unit 
sphere ${\cal S}$. Following [4] and [5], in addition to the 
fall-off conditions we impose the following {\it global parity 
conditions} on the leading terms of the metric and extrinsic 
curvature: 

$$
q_{\bi\bj}{}^{(k)}(-{x^{\bk}\over r})=q_{\bi\bj}{}^{(k)}({x^{\bk}
   \over r}),\hskip 20pt
\chi_{\bi\bj}{}^{(l)}(-{x^{\bk}\over r})=-\chi_{\bi\bj}{}^{(l)}
  ({x^{\bk}\over r}); \eqno(2.3.2a,b)
$$
\ni
i.e. $q_{\bi\bj}{}^{(k)}$ is of even, while $\chi_{\bi\bj}{}^{(l)}$ 
is of odd parity. Furthermore, we assume that the `rests' 
$m_{ab}:=q_{ab}-{}_{0}q_{ab}-r^{-k}q_{ab}{}^{(k)}$ and $k_{ab}:=
\chi_{ab}-r^{-l}\chi_{ab}{}^{(l)}$ satisfy the additional conditions 

$$\eqalignno{
{}_0D_cm_{ab}&=o(r^{-k-1}), \hskip 20pt 
 {}_0D_d{}_0D_cm_{ab}=o(r^{-k-2}), \hskip 20pt
 {}_0D_e{}_0D_d{}_0D_cm_{ab}=o(r^{-k-3}), ... &(2.3.3a)\cr
{}_0D_ck_{ab}&=o(r^{-l-1}), \hskip 20pt
 {}_0D_d{}_0D_ck_{ab}=o(r^{-l-2}), ... &(2.3.3b)\cr}
$$
\ni
which imply that ${}_0D_{e_s}...{}_0D_{e_1}q_{ab}=O(r^{-(k+s)})$, 
$s=1,2,...$, and, similarly, ${}_0D_{e_s}...{}_0D_{e_1}\chi_{ab}=
O(r^{-(l+s)})$. These properties make the calculations easier. The 
parity of these derivatives is $(-)^s$ and $(-)^{s+1}$, respectively. 
The properties $m_{ab}=o(r^{-k})$, ${}_0D_cm_{ab}=o(r^{-k-1})$, 
${}_0D_d{}_0D_cm_{ab}=o(r^{-k-2})$, ..., ${}_0D_{e_s}...{}_0D_{e_1}
m_{ab}=o(r^{-(k+s)})$ of the rest $m_{ab}$ will be denoted by $m_{ab}
=o^s(r^{-k})$, and, similarly, $k_{ab}=o^s(r^{-l})$. Although in the 
actual calculations we will use these (essentially technical) 
assumptions for some finite value of $s$, for the sake of simplicity 
we assume that $m_{ab}=o^\infty(r^{-k})$ and $k_{ab}=o^\infty(r^{-l})$.
We may call the asymptotic end $(\Sigma,q_{ab},\chi_{ab})$ to be 
$(k,l)$-asymptotically flat if for some background metric ${}_0q
_{ab}$ the conditions (2.3.1)-(2.3.3) are satisfied. However, we will 
see in subsection 4.2 that this notion depends on the background 
metric too, thus $(\Sigma,q_{ab},\chi_{ab})$ will be called {\it 
$(k,l)$-asymptotically flat with respect to the background metric 
${}_0q_{ab}$} if (2.3.1)-(2.3.3) are satisfied.

Now let us suppose that $\mu$, $j^a$ and $\sigma^{ab}$ satisfy the 
fall-off and parity conditions (2.2.3) with respect to the Cartesian 
coordinates $x^{\bk}$ on $\Sigma$ with the additional (technical) 
requirement that the `rests' are of order $o^\infty(r^{-m})$. Next 
ask what conditions should we impose to ensure the existence of the 
limit 

$$
Q_m[M,M^a]:=\lim_{r\rightarrow\infty}\int_{B_r}\Bigl(\mu M+j^aM_a
\Bigr){\rm d}\Sigma \eqno(2.3.4)
$$
\ni 
and, considering $\Sigma$ to be a leave of the foliation defined by 
a lapse function $N$, its time derivative 

$$\eqalign{
\dot Q_m[M,M^a]=\lim_{r\rightarrow\infty}\int_{B_r}\Bigl\{&\mu\bigl(
 \dot M-N^aD_aM+M^aD_aN\bigr)+\cr
+&j^a\bigl(\dot M_a-MD_aN+ND_aM-\L_{\bf N}M_a-2\chi_{ab}M^bN\bigr)+\cr
+&\sigma^{ab}N\bigl(M\chi_{ab}+D_{(a}M_{b)}\bigr)+\cr
+&D_a\Bigl(\bigl(\mu M+j^bM_b\bigr)N^a-\bigl(j^aM+\sigma^{ab}M_b\bigr)
 N\Bigr)\Bigr\}{\rm d}\Sigma\cr}\eqno(2.3.5)
$$
\ni
with respect to the time axis $\xi^a:=Nt^a+N^a$. To obtain (2.3.5) we 
used (2.1.2)-(2.1.4). Note that if $K^a=Mt^a+M^a$, then $Q_m[M,M^a]$ 
is just the $n+1$ form of $Q_m[K^a]:=\int_\Sigma K_aT^{ab}t_b{\rm d}
\Sigma$, and $\dot Q_m[M,M^a]$ is the $n+1$ form of $\int_\Sigma(T
^{ab}(\nabla_aK_b)N+D_a((P^a_bt_c-P^a_ct_b)\xi^bT^{cd}K_d)){\rm d}
\Sigma$. 
Thus let us suppose that $M$ and $M_{\bi}$, where the latter is 
defined by $M_a=:M_{\bi}D_ax^{\bi}$, have the asymptotic 
form\footnote{${}^2$}{To be consistent with our previous notations, 
we would have to write $M^{(-A)}$ instead of $M^{(A)}$. However, we 
resolve this apparent inconsistency with the convention that the 
power of $r$ is always a capital, while the power of $1/r$ is always 
a lower case letter. In particular, the leading term in the expansion 
of $f(r,{x^{\bk}\over r})$ can be written as $r^Af^{(A)}({x^{\bk}
\over r})$ or $r^{-a}f^{(a)}({x^{\bk}\over r})$.}

$$\eqalignno{
M(t,x^{\bk})&=r^A\,M^{(A)}(t,{x^{\bk}\over r})+o^\infty( r^A),
 &(2.3.6a)\cr
M_{\bi}(t,x^{\bk})&=r^B\,M^{(B)}_{\bi}(t,{x^{\bk}\over r})+o^\infty(
 r^B)&(2.3.6b)\cr}
$$
\ni
for some $A$, $B$. Substituting these into (2.3.4) we obtain that 
$Q_m[M,M^a]$ exists precisely if 

$$
A\leq1, \hskip12pt B\leq1; \eqno(2.3.7)
$$
\ni
and if the equality holds in these inequalities, then the leading 
terms, $M^{(A)}(t,{x^{\bk}\over r})$ and $M^{(B)}_{\bi}(t,{x^{\bk}
\over r})$, must be {\it odd parity functions} of ${x^{\bk}\over 
r}$, respectively. The existence of $\dot Q_m[M,M^a]$ restricts $N$ 
and $N^a$. Before discussing these restrictions, clarify first the 
conditions for $N$ and $N^a$ by means of which the conservation 
equations (2.1.3) and (2.1.4), and the consequence (2.1.2) of the 
definition of the `time derivative' preserve the fall-off and parity 
conditions (2.2.3) for the matter fields and (2.3.1a-3a) for the 
metric, respectively. Writing $N$ and $N_{\bi}$ in the form of $M$ 
and $M_{\bi}$ given by (2.3.6) with some powers $C$ and $D$, 
substituting them into (2.1.3), (2.1.4) and (2.1.2) and requiring 
their right hand side to have $O(r^{-m})$ order and even parity, 
$O(r^{-m})$ order and even parity and $O(r^{-k})$ order and even 
parity, respectively, we obtain that 

$$\eqalignno{
N(t,x^{\bk})&=r^C\,N^{(C)}(t,{x^{\bk}\over r})+o^\infty(r^C),
 &(2.3.8a)\cr
N_{\bi}(t,x^{\bk})&=2x^{\bk}\rho_{\bk\bi}(t)+\tau_{\bi}(t)+r^F\,\nu
 ^{(F)}_{\bi}(t,{x^{\bk}\over r})+o^\infty(r^F),&(2.3.8b)\cr}
$$
\ni
where $\tau_{\bi}(t)$ and $\rho_{\bi\bj}(t)=-\rho_{\bj\bi}(t)$ are 
independent of the coordinates $\{x^{\bk}\}$ but may be arbitrary 
functions of the coordinate time $t$, the powers $C$ and $F$ satisfy

$$
C\leq{\rm min}\{1,l-k\}, \hskip 20pt F\leq(1-k), \eqno(2.3.9)
$$
\ni
and if the equality holds then the leading terms $N^{(C)}(t,{x^{\bk}
\over r})$ and $\nu^{(F)}_{\bi}(t,{x^{\bk}\over r})$ must be odd 
parity functions of ${x^{\bk}\over r}$, respectively. (The first two 
terms on the right of (2.3.8b) together is just the kernel of the 
Killing operator ${}_0D_{({\bi}}N_{{\bj})}=0$ in (2.1.2).) Note, 
first, that by (2.3.9) $F<1$, because we assumed that $k>0$, and, 
second, that there is no reason to keep the term $\tau_{\bi}(t)$ in 
(2.3.8b) if $F>0$. Substituting (2.3.6) and (2.3.8) into (2.3.5) 
and taking into account the conditions (2.3.7) and (2.3.9), one 
can check that the integral exists (and, in particular, the total 
divergence gives zero). If $K^a=Mt^a+M^a$ is a spacetime Killing 
vector then, by (2.1.10-12), $\dot Q_m[K^a]$ is zero, as it must be 
since then $T^{ab}K_b$ is divergence-free. 

The notion of the $(k,l)$-asymptotic flatness is referring only to 
the asymptotic end of a single spacelike hypersurface $\Sigma$. To 
ensure that the spacetime itself, i.e. the evolution of $\Sigma$, is 
also asymptotically flat, we must ensure the compatibility of the 
boundary conditions with the evolution equations for the geometry. 
Thus let us consider the evolution equation (2.1.7) for the extrinsic 
curvature and ask under what {\it additional} conditions for $N$ and 
$N_a$ do these equations preserve the fall-off and parity conditions 
(2.3.1b-3b) for the extrinsic curvature. An analysis similar to that 
we did above yields that (2.1.7) preserves these asymptotic 
conditions precisely when 

$$\eqalignno{
N(t,x^{\bk})&=2x^{\bk}\beta_{\bk}(t)+\tau(t)+r^E\nu^{(E)}(t,{x^{\bk}
 \over r})+o^\infty(r^E), &(2.3.10a)\cr
N_{\bi}(t,x^{\bk})&=2x^{\bk}\rho_{\bk\bi}(t)+\tau_{\bi}(t)+r^F\nu
 _{\bi}{}^{(F)}(t,{x^{\bk}\over r})+o^\infty(r^F), &(2.3.10b)\cr}
$$
\ni
where $\tau(t)$ and $\beta_{\bi}(t)$ are independent of $\{x^{\bk}
\}$ but may depend on $t$, the powers $E$ and $F$ satisfy 

$$
E\leq(2-l), \hskip 20pt F\leq(1-k), \eqno(2.3.11)
$$
\ni
and if the equality holds in (2.3.11) then $\nu^{(E)}({x^{\bk}\over 
r})$ and $\nu_{\bi}{}^{(F)}({x^{\bk}\over r})$ are of odd parity, 
respectively. Then by (2.3.9) $E\leq C\leq (l-k)$, implying that $E
\leq(1-{1\over2}k)<1$ and $k+E\leq l\leq 2-E$. In addition, if $\tau
(t)\not=0$ then $l<k+2$, and, by (2.3.9), $k\leq l$ even for $E\leq0$. 
If $\beta_{\bi}(t)\not=0$ then $l=k+1$, and hence $E\leq(1-k)<1$. In the 
presence of matter $\tau(t)\not=0$ also implies $l<w$ and $\beta_{\bi}
(t)\not=0$ also implies $l\leq w-1$, where $w$ is the actual order of 
the leading term of the spatial stress (and of the energy density 
and momentum density), which, as we saw, must satisfy $w\geq m$. 
(Interestingly enough, it is just the fall-off conditions $k>0$, 
$l=k+1$ that ensure the existence of the spinor Chern--Simons 
functional $Y$ on spacelike hypersurfaces in 3+1 dimensional 
spacetimes [26], by means of which the vacuum Einstein equations 
can be recovered as the necessary and sufficient condition of the 
invariance of $Y$ with respect to infinitesimal spacetime conformal 
rescalings [23].) For $n=3$ and the {\it a priori} powers $l=k+1=2$ 
the expression (2.3.10) is almost the condition of Beig and \'O 
Murchadha obtained from the investigation of the vacuum evolution 
equations. The only (and, as we will see, important) difference is 
that the evolution equations, even in the presence of matter, allow 
the time dependence of the coefficients in $N$ and $N^a$. 

\bigskip
\bigskip

\ni
{\bf 2.4 Allowed time axes and the asymptotic Killing vectors}\par
\bigskip
\ni
In the previous subsections the time axis $\xi^a$ with respect to 
which the time evolution was defined and the generator $K^a$ of the 
physical quantity in $Q_m[K^a]$ were treated separately: 
While the role of $\xi^a$ was to provide a differential topological 
background, e.g. a foliation of the spacetime and a shift vector to 
carry out the analysis (general time axis), the nature of $K^a$ told 
us whether the corresponding $Q_m[K^a]$ should be interpreted e.g. as 
the energy or a component of the spatial angular momentum of the 
matter fields. Indeed, the lapse and shift parts of $K^a$ were defined 
with respect to the foliation that $\xi^a$ defined. 

However, the structure (2.3.10-11) of $N$ and $N^a$ is compatible with 
(2.3.6-7), or, in other words, the time axes $\xi^a$ can be considered 
as special generators $K^a$. (Note that the vector fields $\xi^a$ and 
$K^a$ obtained here might be tangent to $\Sigma$ or even vanishing, 
and their lapse part might be positive on some subset and negative on 
other subsets of $\Sigma$. In spite of this fact we call $\xi^a$ an 
`allowed time axis'.) 
Furthermore, these two roles are mixed in the Hamiltonian formulation 
of the dynamics of the matter+gravity systems: $M$ and $M^a$ in the 
total Hamiltonian (in particular in its matter part $Q_m[M,M^a]$) play 
the role of the n+1 form of the generator $K^a$, and, at the same time, 
the Hamiltonian generates the time evolution of the states with respect 
to the spacetime vector field $\xi^a=Mt^a+M^a$. Thus we assume that all 
the generators $M$, $M_{\bi}$ and $\bar M$, $\bar M_{\bi}$ also have 
the structure (2.3.10-11) of the allowed time axes and we write

$$\eqalign{
M&=2x^{\bk}B_{\bk}(t)+T(t)+r^E\mu^{(E)}+o^\infty\bigl(r^E\bigr),
 \hskip 20pt 
M_{\bi}=2x^{\bk}R_{\bk\bi}(t)+T_{\bi}(t)+r^F\mu_{\bi}{}^{(F)}+o
 ^\infty\bigl(r^F\bigr),\cr
\bar M&=2x^{\bk}\bar B_{\bk}(t)+\bar T(t)+r^G\bar\mu^{(G)}+o^\infty
 \bigl(r^G\bigr),\hskip 20pt 
\bar M_{\bi}=2x^{\bk}\bar R_{\bk\bi}(t)+\bar T_{\bi}(t)+r^H\bar\mu
 _{\bi}{}^{(H)}+o^\infty\bigl(r^H\bigr),\cr
N&=2x^{\bk}\beta_{\bk}(t)+\tau(t)+r^K\nu^{(K)}+o^\infty\bigl(r^K
 \bigr), \hskip 20pt 
N_{\bi}=2x^{\bk}\rho_{\bk\bi}(t)+\tau_{\bi}(t)+r^L\nu_{\bi}{}^{(L)}+
 o^\infty\bigl(r^L\bigr),\cr}\eqno(2.4.1)
$$
\ni
where $E,F,G,H,K,L\leq(1-k)$ and, in the case of equality here, the 
corresponding coefficient has odd parity. The space of the pairs 
$(M,M^a)$ on $\Sigma$ given by (2.4.1) will be denoted by ${\cal A}$. 
Considering $M$ to be the lapse of a (maybe degenerate) foliation 
of a neighbourhood of $\Sigma$ in the spacetime, ${\cal A}$ can also 
be interpreted as the space of the {\it spacetime vector fields} $\xi
^a:=Mt^a+M^a$, where $t^a$ is the future pointing unit normal to the 
leaves of the foliation. 

According to the double role of the components given by (2.4.1), we 
form another space of spacetime vector fields. Namely, if a lapse $N$ 
is given on $\Sigma$, then let ${\cal A}_N$ be the space of those 
spacetime vector fields $K^a:=Mt^a+M^a$ that are defined with respect 
to the (maybe degenerate) foliation determined by $\Sigma$ and $N$. 
(If $N$ has zeros, then not every element $(M,M^a)$ of ${\cal A}$ 
determines a spacetime vector field in ${\cal A}_N$: if the leaves 
$\Sigma_t$ and $\Sigma_{t'}$ of the foliation intersect each other at 
some point $p$, then only those elements $(M,M^a)$ of ${\cal A}$ can 
determine spacetime vector fields in ${\cal A}_N$, for which $M(t,p)t
^a_p+M^a(t,p)=M(t',p)t'^a_p+M^a(t',p)$, where $t^a_p$ and $t'^a_p$ 
are the future pointing unit normal of $\Sigma_t$ and $\Sigma_{t'}$, 
respectively, at $p$.) 
The structure of the leading two terms e.g. of $M$ and $M_{\bi}$ 
resembles to that of the $n+1$ decomposition of the familiar 
spacetime Killing vectors of the Minkowski spacetime with respect 
to a spacelike hypersurface. However, although the first terms 
are linear in, and the second terms are independent of the spatial 
coordinates, the third terms (which would be analogous to the 
supertranslations of subsection 2.2) may depend on the spatial 
coordinates and may even be diverging. Moreover, although by 
(2.1.12) $P^c_aP^d_b\nabla_{(c}K_{d)}=O(r^{-k})$, i.e. tends to zero 
at infinity, in general neither $P^b_at^c\nabla_{(b}K_{c)}$ nor $t^a
t^b\nabla_{(a}K_{b)}$ tend to zero. 
The space ${\cal A}_N$ does not form a Lie algebra. In fact, if $K^a
:=Mt^a+M^a$ and $\bar K^a:=\bar Mt^a+\bar M^a$ are any two elements 
of ${\cal A}_N$, then the first two terms of their Lie bracket in 
(2.1.15) are given by (2.1.10) 
and (2.1.11), respectively. Substituting (2.4.1) into these formulae 
we find that their asymptotic structure is dominated by $N^{-1}x^{\bm}
x^{\bn}$, which deviates from that of the allowed time axes. 
Moreover, while the time dependence of the Killing vectors in a 
coordinate system adapted to the translations is very specific, the 
time dependence of $M$ and $M_{\bi}$ is not specified at all. In this 
subsection we use this freedom to specify a class of spacetime vector 
fields, whose elements can naturally be interpreted as asymptotic 
Killing vectors. The components of the spacetime vector fields that 
are only allowed time axes will be denoted by Greek letters, as those 
of $N$ and $N^a$ in (2.4.1).

We noted in subsection 2.1 that the parts $P^b_at^c\nabla_{(b}K_{c)}$ 
and $t^at^b\nabla_{(a}K_{b)}$ of the Killing operator acting on some 
$K^a\in{\cal A}_N$ can be required to be zero. Thus let us define the 
allowed time axis $K^a=Mt^a+M^a$ to be {\it a strong asymptotic 
Killing vector} with respect to the foliation characterized by the 
lapse $N$ if its components $M$ and $M^a$ satisfy (2.1.13) and 
(2.1.14). However, to ensure that the components of these asymptotic 
Killing vectors be well defined, a choice for the shift vector $N^a$ 
must also be made. The set of these asymptotic Killing vector fields 
will be denoted by ${\cal A}^0_\xi$, where $\xi^a:=Nt^a+N^a$, which 
will be assumed to be an allowed time axis. 
Obviously, this notion of asymptotic Killing vectors depends sharply 
on the lapse $N$ and the hypersurface $\Sigma$ (on which the foliation 
is based): if a neighbourhood of $\Sigma$ in the spacetime is foliated 
by another lapse $\bar N$ from (maybe) another hypersurface $\bar
\Sigma$ with normal $\bar t^a$, then the corresponding parts $\bar t
^b\bar t^c\nabla_{(b}K_{c)}$ and $\bar P^b_a\bar t^c\nabla_{(b}K_{c)}$ 
of the Killing operator will not be zero. 
(Perhaps, the notation ${\cal A}^0_\xi$ is not very fortunate, and it 
would have to be denoted by ${\cal A}^0_{(\Sigma,N);N^a}$ to stress 
that this notion of asymptotic Killing vectors depends both on $\Sigma$ 
and $N$, and the asymptotic Killing vectors themselves are 
parameterized by using the shift $N^a$ too.) 
Therefore, this notion of the asymptotic Killing vectors appears to 
be unnecessarily strong, and it could be enough to require that the 
parts $P^b_at^c\nabla_{(b}K_{c)}$ and $t^at^b\nabla_{(a}K_{b)}$ of the 
Killing operator be at most of order $O(r^{-k})$ asymptotically, i.e. 
explicitly 

$$\eqalignno{
t^at^b\nabla_{(a}K_{b)}&={1\over N}\Bigl(\dot M+M^aD_aN-N^aD_aM\Bigr)
  =r^P\kappa^{(P)}\bigl(t,{x^{\bk}\over r}\bigr)+o^\infty\bigl(r^P\bigr),
  &(2.4.2)\cr
2P^b_at^c\nabla_{(b}K_{c)}&={1\over N}\Bigl(\dot M_a+\bigl(ND_aM-MD_a
  N\bigr)-2N\chi_{ab}M^b-\L_{\bf N}M_a\Bigr)=\cr
&=\Bigl(r^Q\kappa_{\bi}^{(Q)}\bigl(t,{x^{\bk}\over r}\bigr)+o^\infty
  \bigl(r^Q\bigr)\Bigr)D_ax^{\bi},  &(2.4.3)\cr}
$$
\ni
for some $P,Q\leq-k$ and if $P$ and $Q$ are equal to $-k$ then $\kappa
^{(P)}$ and $\kappa^{(Q)}_{\bi}$ have even parity, respectively. Note 
that (2.4.2) and (2.4.3) can always be solved for $M$ and $M_a$ for 
any given functions $\kappa^{(P)}(t,{x^{\bk}\over r})$ and $\kappa
_{\bi}^{(Q)}(t,{x^{\bk}\over r})$. (2.4.2) and (2.4.3) will be called 
the asymptotic Killing equations. We call the vector field $K^a\in
{\cal A}_N$ {\it asymptotic Killing vector} with respect to the 
foliation determined by the lapse $N$ if its components $M$ and $M^a$ 
are solutions of the asymptotic Killing equations. Clearly, the notion 
of the asymptotic Killing vectors is less sensitive to the deformation 
of $\Sigma$ and $N$ than that of the strong asymptotic Killing 
vectors, but it is still not independent of the foliation that $N$ and 
$\Sigma$ define. 
In particular, as we will see below, the asymptotic Killing vectors 
defined with respect to a foliation for which the lapse $N$ tends to 
zero are different from those defined with respect to ones for which 
$N\rightarrow1$ as $r\rightarrow\infty$. The set of the asymptotic 
Killing vector fields will be denoted by ${\cal A}^K_\xi$. Obviously, 
${\cal A}^0_\xi\subset{\cal A}^K_\xi\subset{\cal A}_N$ for any allowed 
time axis $\xi^a$, and ${\cal A}_N$ can be injected into ${\cal A}$. 

Substituting (2.4.1) into the asymptotic Killing equations (2.4.2) 
and (2.4.3), we obtain 

$$\eqalignno{
\dot B_{\bi}&=-2\Bigl(R_{\bi\bj}\beta^{\bj}-\rho_{\bi\bj}B^{\bj}
 \Bigr),&(2.4.4)\cr
\dot R_{\bi\bj}&=2\Bigl(B_{\bi}\beta_{\bj}-\beta_{\bi}B_{\bj}\Bigr)-
 2\Bigl(R_{\bi\bk}\rho^{\bk}{}_{\bj}-\rho_{\bi\bk}R^{\bk}{}_{\bj}
 \Bigr),&(2.4.5)\cr}
$$
\ni
and, if $E,F\leq0$, we also have 

$$\eqalignno{
\dot T&=-2\Bigl(T_{\bi}\beta^{\bi}-\tau_{\bi}B^{\bi}\Bigr),&(2.4.6)\cr
\dot T_{\bi}&=2\Bigl(T\beta_{\bi}-\tau B_{\bi}\Bigr)-2\Bigl(T^{\bj}
 \rho_{\bj\bi}-\tau^{\bj}R_{\bj\bi}\Bigr),&(2.4.7)\cr}
$$
\ni
independently of $\kappa^{(P)}$ and $\kappa^{(Q)}_{\bi}$. Thus the 
coefficients $B_{\bi}$, $R_{\bi\bj}$, $T$ and $T_{\bi}$ in both of the 
elements of ${\cal A}^0_\xi$ and ${\cal A}^K_\xi$ satisfy (2.4.4-7). 
To calculate the Lie bracket of two asymptotic Killing vectors 
$K^a,\bar K^a\in{\cal A}^K_\xi$ explicitly, we should compute only 
the last three terms on the right hand side of (2.1.15), because, by 
(2.4.2) and (2.4.3), the first two are at most of order $O(r^{1-k})$ 
in general, and zero if $K^a,\bar K^a\in{\cal A}^0_\xi$. The last 
three terms are 

$$\eqalignno{
M^aD_a\bar M&-\bar M^aD_aM=4x^{\bk}\Bigl(R_{\bk\bi}\bar B^{\bi}-\bar 
  R_{\bk\bi}B^{\bi}\Bigr)+2\Bigl(T_{\bi}\bar B^{\bi}-\bar T_{\bi}B
  ^{\bi}\Bigr)+\cr
&+4r^{1-k}{x^{\bk}\over r}\Bigl(\bar R_{\bk\bi}B_{\bj}-R_{\bk\bi}\bar 
  B_{\bj}\Bigr)q^{(k)}{}^{\bi\bj}+o^\infty\bigl(r^{1-k}\bigr),&(2.4.8)
  \cr
\Bigl[M,\bar M\Bigr]^a&D_ax^{\bi}=4x^{\bk}\Bigl(R_{\bk\bj}\bar R^{\bj
  \bi}-\bar R_{\bk\bj}R^{\bj\bi}\Bigr)+2\Bigl(T^{\bj}\bar R_{\bj}{}
  ^{\bi}-\bar T^{\bj}R_{\bj}{}^{\bi}\Bigr)+\cr
&+4r^{1-k}{x^{\bk}\over r}\Bigl({x^{\bj}\over r}\bigl(\bar R_{\bk}{}
  ^{\bm}R_{\bj\bn}-R_{\bk}{}^{\bm}\bar R_{\bj\bn}\bigr)\bigl(\bar
  \partial_{\bm}q^{(k)}{}^{\bn\bi}\bigr)-\cr
&\hskip 20pt -\bigl(R_{\bk\bj}\bar R_{\bm}{}^{\bi}-\bar R_{\bk\bj}
  R_{\bm}{}^{\bi}\bigr)q^{(k)}{}^{\bj\bm}-\bigl(R_{\bk\bj}\bar R^{\bj}
  {}_{\bn}-\bar R_{\bk\bj}R^{\bj}{}_{\bn}\bigr)q^{(k)}{}^{\bn\bi}
  \Bigr)+o^\infty\bigl(r^{1-k}\bigr),&(2.4.9) \cr
\Bigl(\bar MD^aM&-MD^a\bar M\Bigr)D_ax^{\bi}=4x^{\bk}\Bigl(\bar B
  _{\bk}B^{\bi}-B_{\bk}\bar B^{\bi}\Bigr)+2\Bigl(\bar TB^{\bi}-T\bar 
  B^{\bi}\Bigr)+\cr
&+2r^G\Bigl(\bar\mu^{(G)}B^{\bi}-{x^{\bk}\over r}B_{\bk}\bigl(G\bar 
  \mu^{(G)}{x^{\bi}\over r}+\bigl(\bar\partial_{\bj}\bar\mu^{(G)}
  \bigr)\bigl({}_0q^{\bj\bi}-{x^{\bj}x^{\bi}\over r^2}\bigr)\bigr)
  \Bigr)+o^\infty\bigl(r^G\bigr)-\cr
&-2r^E\Bigl(\mu^{(E)}\bar B^{\bi}-{x^{\bk}\over r}\bar B_{\bk}\bigl(E
  \mu^{(E)}{x^{\bi}\over r}+\bigl(\bar\partial_{\bj}\mu^{(E)}\bigr)
  \bigl({}_0q^{\bj\bi}-{x^{\bj}x^{\bi}\over r^2}\bigr)\bigr)\Bigr)+
  o^\infty\bigl(r^E\bigr),&(2.4.10)\cr}
$$
\ni
where e.g. $\bar\partial_{\bj}\mu^{(E)}$ denotes the partial 
derivative of $\mu^{(E)}(t,{x^{\bk}\over r})$ with respect to its 
argument ${}_0v^{\bj}={x^{\bj}\over r}$, $q^{(k)}{}^{\bi\bj}:={}_0q
^{\bi\bk}\,{}_0q^{\bj\bl}q^{(k)}_{\bk\bl}$, and we used that $E,F,G,
H\leq(1-k)$. The right hand side of (2.4.8-10) have the form of 
(the components of) an asymptotic Killing vector, and, as a simple 
calculation shows, the coefficients in $\tilde K^a:=[K,\bar K]^a$, 
given explicitly by 

$$\eqalignno{
\tilde B_{\bi}&=2\bigl(R_{\bi\bj}\bar B^{\bj}-\bar R_{\bi\bj}B^{\bj}
 \bigr),&(2.4.11)\cr
\tilde R_{\bi\bj}&=2\bigl(R_{\bi\bk}\bar R^{\bk}{}_{\bj}-\bar R_{\bi
 \bk}R^{\bk}{}_{\bj}+\bar B_{\bi}B_{\bj}-B_{\bi}\bar B_{\bj}\bigr),
 &(2.4.12)\cr
\tilde T_{\bi}&=2\bigl(T^{\bj}\bar R_{\bj\bi}-\bar T^{\bj}R_{\bj\bi}+
 \bar TB_{\bi}-T\bar B_{\bi}\bigr),&(2.4.11)\cr
\tilde T&=2\bigl(T_{\bi}\bar B^{\bi}-\bar T_{\bi}B^{\bi}\bigr),
 &(2.4.14)\cr}
$$ 
\ni
also satisfy (2.4.4-7). However, in general neither ${\cal A}^K_\xi$ 
nor ${\cal A}^0_\xi$ form a Lie algebra with respect to the spacetime 
Lie bracket. In fact, for any two spacetime vector fields $K^a$ and 
$\bar K^a$ one has $\L_{[{\bf K},\bar{\bf K}]}g_{ab}=\L_{\bf K}\L
_{\bar{\bf K}}g_{ab}-\L_{\bar{\bf K}}\L_{\bf K}g_{ab}$, and for $K^a,
\bar K^a\in{\cal A}^K_\xi$ one can form the parts $t^at^b\L_{[{\bf K},
\bar{\bf K}]}g_{ab}$, $P^a_ct^b\L_{[{\bf K},\bar{\bf K}]}g_{ab}$ and 
$P^a_cP^b_d\L_{[{\bf K},\bar{\bf K}]}g_{ab}$. Then, using (2.1.1) 
and the asymptotic Killing equations, one can show that these parts 
contain terms like $MN^{-1}$ or $\bar MN^{-1}$ times factors of order 
$O(r^{-k})$. Thus for general lapse function $N$ the order of these 
parts is not $O(r^{-k})$. Similarly, ${\cal A}^0_\xi$ does not close 
to a Lie algebra either. On the other hand, by (2.4.8-10) both ${\cal 
A}^K_\xi$ and ${\cal A}^0_\xi$ are `essentially' Lie algebras. Next 
we clarify in what sense do they form Lie algebras. 

Let ${\cal G}$ denote the set of the special elements $(\nu,\nu^a)$ 
of ${\cal A}$, where 

$$
\nu(t,x^{\bk})=r^M\nu^{(M)}(t,{x^{\bk}\over r})+o^\infty\bigl(r^M
\bigr), \hskip 20pt
\nu_{\bi}(t,x^{\bk})=r^N\nu^{(N)}_{\bi}(t,{x^{\bk}\over r})+o^\infty
\bigl(r^N\bigr), \eqno(2.4.15)
$$
\ni
for some $M,N\leq(1-k)$ and the leading terms have odd parity if 
$M=1-k$ and $N=1-k$, respectively. Repeating the construction above, 
${\cal G}$ can also be considered as the space of the spacetime 
vector fields $\nu t^a+\nu^a$, where $\nu$ is considered to be the 
lapse of a (maybe degenerate) foliation and $t^a$ is the unit normal 
to the leaves of this foliation. 
If a lapse $N$ is given on $\Sigma$ then we can define ${\cal G}_N$ 
in a quite analogous way as we did above, and introduce ${\cal G}^K
_\xi:={\cal A}^K_\xi\cap{\cal G}_N$. Then by (2.4.8-10) ${\cal G}^K
_\xi$ behaves as an ideal in ${\cal A}^K_\xi$: $[K,k]^a\in{\cal G}
^K_\xi$ for any $k^a\in{\cal G}^K_\xi$ and $K^a\in{\cal A}^K_\xi$. 
We will see in subsections 3.2 and 3.3 below that, at least in the 
Hamiltonian framework, the theory's gauge transformations are 
generated by precisely the elements of ${\cal G}$. Hence we call the 
elements of ${\cal G}$ gauge generators. 
Since ${\cal G}^K_\xi\subset{\cal A}^K_\xi$ is a subspace, one may 
form the quotient space ${\cal A}^K_\xi/{\cal G}^K_\xi$. This is 
spanned by the coefficients $B_{\bi}$, $R_{\bi\bj}$, $T_{\bi}$ and 
$T$, and by (2.4.11-14) it can be endowed with a natural Lie algebra 
structure. 
Similarly, although ${\cal A}^0_\xi$ is not closed with respect to 
the Lie bracket, it almost closes: just by (2.4.4-7) and (2.4.11-14) 
the Lie bracket $[K,\bar K]^a$ of any two $K^a,\bar K^a\in{\cal A}^0
_\xi$ deviates from an element of ${\cal A}^0_\xi$ only by an element 
of ${\cal G}_N$. To determine the structure of ${\cal A}^K_\xi/
{\cal G}^K_\xi$ and of ${\cal A}^0_\xi$, we must evaluate (2.4.4-7) 
and (2.4.11-14). 

Applying (2.4.4-7) to $N$ and $N_{\bi}$ themselves too, we obtain 
that $\tau$, $\tau_{\bi}$, $\beta_{\bi}$ and $\rho_{\bi\bj}$ are 
constant, i.e. apart from the gauge generator contents, {\it the 
coefficients of an asymptotic Killing vector with respect to the 
differential topological background defined by itself are time 
independent}. In particular, $\tau=\tau_{\bi}=\beta_{\bi}=\rho_{\bi
\bj}=0$ corresponds to time axes that are pure gauge generators $\xi
^a=\nu t^a+\nu^a$, whenever the components $T$, $T_{\bi}$, $B_{\bi}$ 
and $R_{\bi\bj}$ of $K^a\in{\cal A}^K_\xi$ are all time independent, 
and the corresponding $M$ and $M_{\bi}$ reduce to those given by 
Beig and \'O Murchadha. 
However, we saw at the end of subsection 2.2 that such a time axis 
does not provide an appropriate framework in which even the familiar 
Killing vectors of the Minkowski spacetime could be recovered. To be 
able to recover them we had to assume that, with the notations of the 
present subsection, $\tau=1$. Thus, if $\tau=1$ and $\tau_{\bi}=\beta
_{\bi}=\rho_{\bi\bj}=0$ then the coefficients $T$, $B_{\bi}$ and 
$R_{\bi\bj}$ are time independent, but $T_{\bi}(t)=T_{\bi}-2tB_{\bi}$, 
where $T_{\bi}$ is constant. Thus, the corresponding asymptotic 
Killing vector $K^a$ has exactly the same structure as that of the 
general Killing vector of the Minkowski spacetime in the coordinate 
system based on an asymptotically flat hypersurface approaching the 
Cartesian one. Therefore, the notion of the asymptotic Killing vectors 
does depend on the foliation coming from $\Sigma$ and $N$. Naturally, 
by (2.3.5) $Q_m[K^a]$ is still not conserved for general asymptotic 
Killing vectors, but, as we will see in subsection 4.1, the total 
energy-momentum and angular momentum of the matter+gravity system are 
already conserved. 

If $E,F,...>0$ then the translation generators $T$ and $T_{\bi}$ in 
(2.4.1) cannot be isolated from the diverging gauge generators, and 
hence by (2.4.11-12) the quotient space ${\cal A}^K_\xi/{\cal G}^K
_\xi$ endowed with the Lie product (2.4.11-14) is isomorphic to the 
Lorentz Lie algebra $so(1,n)$. If, however, $E,F,...\leq0$ then $T$ 
and $T_{\bi}$ are well defined in (2.4.1). In particular, if $E,F,...
<0$, then asymptotically $T$ and $T_{\bi}$ dominate the 
(asymptotically vanishing) gauge generators. 
For $E,F,...=0$ the gauge generators do not vanish asymptotically 
(which therefore can be interpreted as {\it supertranslations}), but 
in the limiting case $E,F, ...=(1-k)=0$ the gauge generators have 
{\it odd parity}, while $T$ and $T_{\bi}$, being independent of the 
spatial coordinates, have {\it even parity}. Thus the odd parity 
supertranslations are proper gauge generators, while the even parity 
ones, which are called the translations, belong to ${\cal A}^K_\xi-
{\cal G}^K_\xi$. Thus, by (2.4.11-14), ${\cal A}^K_\xi/{\cal G}^K
_\xi$ is isomorphic to the Poincare Lie algebra. Therefore, {\it the 
algebra of the asymptotic Killing vectors modulo gauge generators 
depends on the fall-off of the metric: for slow ($0<k<1$) fall-off 
we have only the Lorentz Lie algebra, but for faster ($k\geq1$) 
fall-off translations emerge naturally and we have a Poincare 
structure for ${\cal A}^K_\xi/{\cal G}^K_\xi$.} 
This result should be intuitively obvious: if the asymptotic end 
is asymptotically flat in any sense then it is becoming spherical 
asymptotically and hence the rotation group (and its relativistic 
extension, the Lorentz group) emerges naturally, but the 
displacements of the centre of the asymptotic rotations become 
asymptotic symmetries, i.e. the asymptotic translations emerge 
and hence the symmetry group is the Poincare group, only if the 
geometry falls-off rapidly enough. Finally, we note that the fact 
that the coefficients $B_{\bi}$, $R_{\bi\bj}$, $T_{\bi}$ and $T$ 
satisfy (2.4.4-7) independently of whether $K^a$ belongs to ${\cal 
A}^K_\xi$ or ${\cal A}^0_\xi$ implies that the factor spaces ${\cal 
A}^K_\xi/{\cal G}^K_\xi$ and ${\cal A}^0_\xi/{\cal G}^0_\xi$ are 
isomorphic, where ${\cal G}^0_\xi:={\cal G}_N\cap{\cal A}^0_\xi$, 
and hence ${\cal A}^0_\xi/{\cal G}^0_\xi$ also has the Lie algebra 
structure that ${\cal A}^K_\xi/{\cal G}^K_\xi$ does.

\bigskip
\bigskip

\ni
{\lbf 3. The Hamiltonian phase space of the vacuum GR}\par
\medskip
\ni
{\bf 3.1 The (partially reduced) phase space and the constraints}\par
\medskip
\ni
Based on the $m=n+1$ decomposition, by the {\it a priori} configuration 
variables we would have to mean the fields $N$, $N^a$ and $q_{ab}$ on 
a connected $n$ dimensional manifold $\Sigma$ of subsection 2.1 (see 
also [27]). As is well known, however, by carrying out the full 
Hamiltonian analysis with these variables, the fields $N$ and $N^a$ 
would turn out to be pure gauge variables, and the {\it a priori} 
Hamiltonian phase space could be partially reduced to the cotangent 
bundle $T^*{\cal Q}$ of the (partially reduced) configuration space 
${\cal Q}:=\{\,q_{ab}\,+\,{\rm boundary\,\,conditions}\,\}$. Thus our 
analysis will be based on the partially reduced configuration space 
${\cal Q}$ and its cotangent bundle $T^*{\cal Q}$. Let $q_{ab}(u)$, 
$u\in(-\epsilon,\epsilon)$, be any smooth 1-parameter family of 
metrics on $\Sigma$ from ${\cal Q}$ for some $\epsilon>0$. Then we 
define $\delta q_{ab}:=({\rm d}q_{ab}(u)/{\rm d}u)_{u=0}$, which is 
the tangent vector of the curve $q_{ab}(u)$ at $q_{ab}:=q_{ab}(0)$, 
i.e. $\delta q_{ab}\in T_{q_{ab}}{\cal Q}$. Obviously, $\delta q
_{ab}$ satisfies the same boundary conditions that $q_{ab}$ does. 
The elements of $T^*{\cal Q}$ are the pairs $(q_{ab},\tilde p^{ab})$, 
where 

$$
\tilde p^{ab}: T_{q_{ab}}{\cal Q}\rightarrow{\bf R}:\delta q_{ab}
\mapsto\langle\,\tilde p^{ab},\delta q_{ab}\,\rangle:=\int_\Sigma
\tilde p^{ab}\delta q_{ab}{\rm d}^nx.  \eqno(3.1.1)
$$
\ni
Thus the canonical momentum $\tilde p^{ab}$ is a contravariant 
symmetric tensor density of weight one on $\Sigma$, which is a 1-form 
on ${\cal Q}$, and the requirement of the finiteness of its action on 
the tangent vectors $\delta q_{ab}$ gives boundary conditions for 
$\tilde p^{ab}$. The symplectic 2-form on $T^*{\cal Q}$ is the 
canonical one: for any two tangent vectors $(\delta q_{ab},\delta
\tilde p^{ab}),(\delta'q_{ab},\delta'\tilde p^{ab})\in T_{(q_{ab},
\tilde p^{ab})}(T^*{\cal Q})$ the value of the symplectic 2-form 
$\Omega_{(q_{ab},\tilde p^{ab})}$ is 

$$\eqalign{
\Omega_{(q_{ab},\tilde p^{ab})}:&T_{(q_{ab},\tilde p^{ab})}(T^*{\cal 
 Q})\times T_{(q_{ab},\tilde p^{ab})}(T^*{\cal Q})\rightarrow{\bf R}\cr
:&\Bigl(\bigl(\delta q_{ab},\delta\tilde p^{ab}\bigr),\bigl(\delta'q
 _{ab},\delta'\tilde p^{ab}\bigr)\Bigr)\mapsto \int_\Sigma\Bigl(\delta
 \tilde p^{ab}\delta'q_{ab}-\delta'\tilde p^{ab}\delta q_{ab}\Bigr)
 {\rm d}^nx.\cr}
\eqno(3.1.2)
$$
\ni
In terms of the metric and extrinsic curvature the canonical momentum 
is well known to be 

$$
\tilde p^{ab}={1\over2\kappa}\sqrt{\vert q\vert}\bigl(\chi^{ab}-\chi q
^{ab}\bigr).\eqno(3.1.3)
$$
\ni
Thus the extrinsic curvature (i.e. by (2.1.2) the velocity $\dot q
_{ab}$) can be expressed by the momenta: $\sqrt{\vert q\vert}\chi
^{ab}=2\kappa(\tilde p^{ab}-{1\over(n-1)}\tilde p^{ef}q_{ef}q^{ab})$.

The analysis of the field equations in the previous section lead us 
to the link $l=k+1$ between the $r^{-k}$ and $r^{-l}$ {\it a priori} 
fall-off of the metric and extrinsic curvature. Thus, via (3.1.3), we 
obtain the fall-off 

$$
\tilde p^{\bi\bj}(x^{\bk})={1\over r^{k+1}}\tilde p^{\bi\bj}{}^{(k+1)}
({x^{\bk}\over r})+o(r^{-(k+1)}) \eqno(3.1.4)
$$
\ni
and the parity condition 

$$
\tilde p^{\bi\bj}{}^{(k+1)}(-{x^{\bk}\over r})=-\tilde p^{\bi\bj}{}
^{(k+1)}({x^{\bk}\over r}), \eqno(3.1.5)
$$
\ni
i.e. $\tilde p^{\bi\bj}{}^{(k+1)}$ is of odd parity. In addition, the 
`rest' $\pi^{ab}:=\tilde p^{ab}-r^{-(k+1)}\tilde p^{ab}{}^{(k+1)}$ 
also satisfies the conditions 

$$
{}_0D_c\pi^{ab}=o(r^{-k-2}), \hskip 20pt
{}_0D_d{}_0D_c\pi^{ab}=o(r^{-k-3}), ... \eqno(3.1.6)
$$
\ni
These imply that ${}_0D_{e_s}...{}_0D_{e_1}\tilde p^{ab}=O(r^{-(k+s+
1)})$, $s=1,2,...$, and the parity of the leading term is $(-)^{s+1}$. 
Following the notations of subsection 2.3, this property of the rest 
will be denoted by $\pi_{ab}=o^\infty(r^{-(k+1)})$. 

By the requirement of the finiteness of (3.1.1) $\tilde p^{ab}$ must 
satisfy 

$$
\oint_{{\cal S}_r}\tilde p^{ab}\delta q_{ab}{\rm d}{\cal S}_r=
o(r^{-1}) \eqno(3.1.7)
$$
\ni
for any $\delta q_{ab}$, which implies that $n\leq k+l=2k+1$, i.e. 
$k\geq{1\over2}(n-1)$. (If we wanted (3.1.7) to be satisfied 
without global integral conditions, and in particular the parity 
condition, then we would have to require $n<k+l$, which turns out 
to be too strong.) Thus the canonical momentum can be interpreted 
geometrically (as the integral kernel of a 1-form on ${\cal Q}$) 
precisely when $k\geq{1\over2}(n-1)$. In particular, if $m=n+1
\geq4$, then $k\geq1$, and hence the Hamiltonian framework already 
excludes the possibility of a slower, e.g. $r^{-{1\over2}}$, 
fall-off, and yields the Poincare structure for ${\cal A}^K_\xi/
{\cal G}^K_\xi$. Slow fall-off is allowed only in 3 spacetime 
dimension. Since $\delta\tilde p^{ab}$ satisfies the same boundary 
conditions that $\tilde p^{ab}$ does, these asymptotic and parity 
conditions ensure that the canonical symplectic 2-form $\Omega$ 
given pointwise by (3.1.2) is also well defined. 

As is well known, although the two {\it vacuum} constraints $c=0$ and 
$c_a=0$, given by (2.1.5) and (2.1.6) in terms of the Lagrangian 
variables $q_{ab}$ and $\dot q_{ab}$, do depend on the lapse $N$ and 
the shift $N^a$, their expressions by the canonical variables $q_{ab}$ 
and $\tilde p^{ab}$, 

$$\eqalignno{
\tilde{\cal C}&:=\tilde c\bigl(N,N^e,q_{ef};\dot q_{ef}(N,N^c,q_{cd},
 \tilde p^{cd})\bigr)=\cr
&\,=-{1\over2\kappa}\sqrt{\vert q\vert}\Bigl(R+{4\kappa^2\over\vert q
 \vert}\bigl({1\over(n-1)}(\tilde p^{ab}q_{ab})^2-\tilde p^{ab}\tilde p
 _{ab}\bigr)\Bigr),&(3.1.8a)\cr
\tilde{\cal C}_a&:=\tilde c_a\bigl(N,N^e,q_{ef};\dot q_{ef}(N,N^c,q
 _{cd},\tilde p^{cd})\bigr)=-2q_{ab}D_c\tilde p^{bc}, &(3.1.8b)\cr}
$$
\ni
are independent of $N$ and $N^a$. Here $\tilde c:=\sqrt{\vert q\vert}
c$ and $\tilde c_a:=\sqrt{\vert q\vert}c_a$, the density weighted 
Lagrangian constraints, and we used the definition $D_e\tau^{a...}
_{b...}:=\vert q\vert^{w\over2}D_eT^{a...}_{b...}$ of the covariant 
derivative of the tensor density $\tau^{a...}_{b...}:=\vert q\vert
^{w\over2}T^{a...}_{b...}$ of weight $w$, where $T^{a...}_{b...}$ is 
a tensor field. Asymptotically

$$
\tilde{\cal C}\bigl(q_{ef},\tilde p^{ef}\bigr)=O\bigl(r^{-(k+2)}\bigr), 
\hskip 20pt
\tilde{\cal C}_a\bigl(q_{ef},\tilde p^{ef}\bigr)=O\bigl(r^{-(k+2)}
\bigr), \eqno(3.1.9a,b)
$$
\ni
and the leading terms are of {\it even} parity. The constraint 
functions $C:T^*{\cal Q}\rightarrow{\bf R}$ defining the constraint 
`surface' $\Gamma$ in $T^*{\cal Q}$ by their vanishing are 

$$
C[\nu,\nu^a]:=\int_\Sigma\Bigl(\tilde{\cal C}\bigl(q_{cd},\tilde p^{cd}
\bigr)\nu+\tilde{\cal C}_a\bigl(q_{cd},\tilde p^{cd}\bigr)\nu^a\Bigr)
{\rm d}^nx. \eqno(3.1.10)
$$
\ni
Here $\nu$ and $\nu^a$ are smearing (test) fields on $\Sigma$. The 
requirement of the finiteness of $C[\nu,\nu^a]$ yields that the 
smearing fields $\nu$ and $\nu_a=:\nu_{\bi}{}_0D_ax^{\bi}$ must 
satisfy 

$$\eqalignno{
\nu\bigl(t,x^{\bk}\bigr)&=r^M\nu{}^{(M)}\bigl(t,{x^{\bk}\over r}
 \bigr)+o^\infty\bigl(r^M\bigr), &(3.1.11a)\cr
\nu_{\bi}\bigl(t,x^{\bk}\bigr)&=r^N\nu_{\bi}{}^{(N)}\bigl(t,{x^{\bk}
 \over r}\bigr)+o^\infty\bigl(r^N\bigr), &(3.1.11b)\cr}
$$
\ni
where $M,N\leq k+2-n=(1-k)+(1+2k-n)$ and if the equality holds in 
these inequalities then $\nu{}^{(M)}$ and $\nu_{\bi}{}^{(N)}$, 
respectively, must be {\it odd} parity functions. Note that, in 
particular, for the slowest possible fall-off of the metric, $k={1
\over2}(n-1)$, the powers $M$ and $N$ would not be greater than 
$(1-k)$. Hence the corresponding spacetime vector field $k^a=\nu 
t^a+\nu^a$ would belong to ${\cal G}$ of subsection 2.4. 

\bigskip

\ni
{\bf 3.2 The constraint algebra}\par
\medskip
\ni
Let $(q_{ab}(u),\tilde p^{ab}(u))$ be a curve in $T^*{\cal Q}$ through 
the point $(q_{ab},\tilde p^{ab})$ with tangent $(\delta q_{ab},\delta
\tilde p^{ab})$. Then the derivative of the constraint function 
$C[\nu,\nu^e]$ in this direction is 

$$\eqalign{
\delta C[\nu,\nu^e]&=\int_\Sigma\Bigl({\delta C[\nu,\nu^e]\over\delta 
 q_{ab}}\delta q_{ab}+{\delta C[\nu,\nu^e]\over\delta\tilde p^{ab}}
 \delta\tilde p^{ab}\Bigr){\rm d}^nx-\cr
&-\int_\Sigma D_e\Bigl({1\over2\kappa}\nu\bigl(q^{ab}D^e\delta q_{ab}
 -q^{ea}D^b\delta q_{ab}\bigr)+{1\over2\kappa}\bigl(q^{ea}D^b\nu-q^{ab}
 D^e\nu\bigr)\delta q_{ab}+\cr
&\hskip 40pt +\bigl(2\nu^a\tilde p^{be}\delta q_{ab}-\nu^e\tilde p^{ab}
 \delta q_{ab}+2\nu_a\delta\tilde p^{ae}\bigr)\Bigr)\sqrt{\vert q\vert}
 {\rm d}^nx,\cr}\eqno(3.2.1)
$$
\ni
where 

$$\eqalignno{
{\delta C[\nu,\nu^e]\over\delta q_{ab}}=&{1\over2\kappa}\sqrt{\vert 
 q\vert}\Bigl\{\nu\Bigl(R^{ab}-Rq^{ab}+{8\kappa^2\over\vert q\vert}
 \bigl(\tilde p^a{}_c\tilde p^{cb}-{1\over(n-1)}q_{cd}\tilde p^{cd}
 \tilde p^{ab}\bigr)\Bigr)+\cr
&+D^aD^b\nu-q^{ab}D_eD^e\nu\Bigr\}-{1\over2}\nu\tilde{\cal C}q^{ab}-
 \L_{\bf \nu}\tilde p^{ab}, &(3.2.2)\cr
{\delta C[\nu,\nu^e]\over\delta\tilde p^{ab}}=&{4\kappa\over\sqrt{
 \vert q\vert}}\nu\Bigl(\tilde p_{ab}-{1\over(n-1)}\tilde p^{cd}q_{cd}
 q_{ab}\Bigr)+\L_{\bf \nu}q_{ab}.&(3.2.3)\cr}
$$
\ni
Here we used the definition $\L_{\bf X}\tau^{a...}_{b...}:=\vert q
\vert^{w\over2}(\L_{\bf X}T^{a...}_{b...}+w({\rm div}\,{\bf X})T
^{a...}_{b...})$ of the Lie derivative of the tensor density $\tau
^{a...}_{b...}:=\vert q\vert^{w\over2}T^{a...}_{b...}$ of weight $w$ 
along the vector field $X^a$, where ${\rm div}\,{\bf X}$ is the 
divergence $D_eX^e$ of $X^a$ with respect to the natural (metric) 
volume form. 
Since we would like to recover e.g. the familiar field equations 
in the Hamiltonian framework in their standard form instead of 
some of their distributional generalizations, we must require the 
functional differentiability of various functions on the phase 
space in the strong sense of [21]. Thus, in particular, the 
boundary terms in (3.2.1) must yield zero. Evaluating the leading 
order and parity of the terms in the total divergence of (3.2.1), 
it is easy to check that the fall-off and parity conditions 
imposed on $\nu$ and $\nu_{\bi}$ in (3.1.11) already imply the 
vanishing of the integral of the total divergence. Thus $C[\nu,\nu
^a]$ are already functionally differentiable. 

The analysis of Beig and \'O Murchadha given in Appendix A of their 
paper [5] shows that the vanishing of the functional derivatives of 
$C[\nu,\nu^a]$ together with the constraints $C[\nu,\nu^a]=0$ 
themselves imply the vanishing of $\nu$ and $\nu^a$. Therefore, the 
constraint `surface' $\Gamma\subset T^*{\cal Q}$ that $C[\nu,\nu^a]
=0$ defines is `nondegenerate'. It might be interesting to note that 
in the closed case (i.e. if $\Sigma$ is compact with no boundary) 
$C[\nu,\nu^a]$ does have critical points on the `surface' $C[\nu,\nu
^a]=0$, and these critical points represent flat spacetimes [23].

The Hamiltonian vector field of a (functionally differentiable) 
function $F:T^*{\cal Q}\rightarrow{\bf R}$ is defined to be the vector 
field $X_F$ on $T^*{\cal Q}$ given explicitly by $X_F=(\delta F/
\delta q_{ab},-\delta F/\delta\tilde p^{ab})$, and the Poisson bracket 
of two differentiable functions, $F$ and $G$, is defined by $\{F,G\}:=
2\Omega(X_F,X_G)=X_F(G)$. Let $\nu$, $\bar\nu$ and $\nu_{\bi}$, $\bar
\nu_{\bi}$ have the structure (3.1.11a) and (3.1.11b), respectively. 
Then the `components' of the Hamiltonian vector field of the 
constraint function $C[\nu,\nu^a]$ are given by (3.2.2) and (3.2.3), 
and the Poisson bracket of the constraint functions $C[0,\nu^a]$ and 
$C[0,\bar\nu^a]$ is 

$$\eqalign{
\Bigl\{C[0,\nu^a]&,C[0,\bar\nu^a]\Bigr\}=-\int_\Sigma\Bigl(\bigl(\L
 _{\bf \nu}q_{ab}\bigr)\bigl(\L_{\bf \bar\nu}\tilde p^{ab}\bigr)-
 \bigl(\L_{\bf \bar\nu}q_{ab}\bigr)\bigl(\L_{\bf \nu}\tilde p^{ab}
 \bigr)\Bigr){\rm d}^nx=\cr
&=\int_\Sigma D_e\Bigl(\nu^e\tilde p^{ab}\L_{\bf \bar\nu}q_{ab}-
 \bar\nu^e\tilde p^{ab}\L_{\bf \nu}q_{ab}-2\tilde p^{ef}[\nu,\bar
 \nu]_f\Bigr){\rm d}^nx-\int_\Sigma \tilde{\cal C}_a[\nu,\bar\nu]^a
 {\rm d}^nx.\cr}\eqno(3.2.4)
$$
\ni
In general, the integral of the total divergence on the right is 
not zero. The condition of its vanishing is $N+\bar N\leq k+3-n$, 
and this also ensures the existence of the second integral and that 
$[\nu,\bar\nu]^a$ has the structure (3.1.11b). The Poisson bracket 
of $C[0,\nu^a]$ and $C[\nu,0]$ is 

$$\eqalign{
\Bigl\{C[0,\nu^a]&,C[\bar\nu,0]\Bigr\}={1\over\kappa}\int_\Sigma D_e
 \Bigl(\bar\nu\bigl(R^e{}_f-{1\over2}R\delta^e_f\bigr)\nu^f+\bigl(D_f
 \nu^e\bigr)\bigl(D^f\bar\nu\bigr)-\bigl(D^e\bar\nu\bigr)\bigl(D_f
 \nu^f\bigr)+\cr
&+{2\kappa^2\over\vert q\vert}\bar\nu\nu^e\bigl(\tilde p^{ab}\tilde 
 p_{ab}-{1\over(n-1)}[\tilde p^{ab}q_{ab}]^2\bigr)\Bigr)\sqrt{\vert q
 \vert}{\rm d}^nx-\int_\Sigma\tilde{\cal C}\bigl(\L_{\bf \nu}\bar\nu
 \bigr){\rm d}^nx.\cr}\eqno(3.2.5)
$$
\ni
The vanishing of the integral of the total divergence can be 
ensured by $N+\bar M\leq3-n$. Furthermore, the condition of the 
finiteness of the second integral is $N+\bar M\leq k+3-n$, which, 
at the same time, ensures that $\nu^aD_a\bar\nu$ has the structure 
(3.1.11a). Finally, the Poisson bracket of $C[\nu,0]$ and $C[\bar
\nu,0]$ is 

$$\eqalign{
\Bigl\{C[\nu,0],C[\bar\nu,0]\Bigr\}&=2\int_\Sigma D_a\bigl(\nu\tilde 
p^{ab}D_b\bar\nu-\bar\nu\tilde p^{ab}D_b\nu\bigr){\rm d}^nx-\int
_\Sigma\tilde{\cal C}_a\bigl(\bar\nu D^a\nu-\nu D^a\bar\nu\bigr){\rm 
d}^nx.\cr}\eqno(3.2.6)
$$
\ni
Here the integral of the total divergence is vanishing if $M+
\bar M\leq k+3-n$, which condition, at the same time, ensures the 
finiteness of the second integral and that $\bar\nu D^a\nu-\nu D^a
\bar\nu$ has the structure (3.1.11b). Thus, to summarize, in 
addition to (3.1.11), the orders of the smearing fields must 
also satisfy $N+\bar N, M+\bar M\leq k+3-n$ and $N+M\leq3-n$. These 
conditions can be satisfied by requiring that the powers $M$ and 
$N$ in (3.1.11) satisfy 

$$
M,N\leq\bigl(1-k\bigr). \eqno(3.2.7)
$$
\ni
By $k\geq{1\over2}(n-1)$ we have $2+k-n=(1-k)+2k+1-n\geq(1-k)$, i.e. 
the condition (3.2.7) is stronger than $M,N\leq2+k-n$, and, in fact, 
$N+M,N+\bar N,M+\bar M\leq2(1-k)\leq(3-n)<3+k-n$. Note that, without 
further restrictions on the power $k$ or the dimension $n$, this is 
the greatest possible bound for $M$ and $N$, because, for the 
allowed smallest value of the rate of the fall-off of the metric, 
$k={1\over2}(n-1)$, $N,M\leq(1-k)=2+k-n$ and $N+M\leq2(1-k)=3-n$. 
Thus, under the condition (3.2.7), the expressions (3.2.4-6) give 
the familiar Lie algebra ${\cal C}$ of the constraint functions with 
the Lie product 

$$
\Bigl\{C[\nu,\nu^a],C[\bar\nu,\bar\nu^a]\Bigr\}=-C\bigl[\L_{\bf \nu}
\bar\nu-\L_{\bf \bar\nu}\nu, [\nu,\bar\nu]^a-\bigl(\nu D^a\bar\nu-
\bar\nu D^a\nu \bigr)\bigr]. \eqno(3.2.8)
$$
\ni
In particular, the Hamiltonian vector fields of the constraint 
functions are tangent to $\Gamma$ on $\Gamma$, i.e. ${\cal C}$ is a 
so-called first class constrained system. Next we clarify how 
this constraint algebra is related to ${\cal G}$. 

The fall-off conditions (3.2.7) show that the smearing fields $\nu$ 
and $\nu^a$ are precisely the elements of ${\cal G}$. Thus, with the 
notation $k^a:=\nu t^a+\nu^a$ in the spacetime picture, we can write 
$C[k^a]:=C[\nu,\nu^a]$, and then $C:{\cal G}\rightarrow{\cal C}:k^a
\mapsto C[k^a]$ is surjective, which is obviously linear. If $C[k^a]$ 
were the zero function in ${\cal C}$ for some $k^a\in{\cal G}$, then 
the corresponding Hamiltonian vector field $X_{C[k^a]}$ would also be 
zero. Thus, by the result of Beig and \'O Murchadha above, $k^a=0$ 
would have to be held, i.e. $C$ is a vector space isomorphism. But in 
general this is not a Lie algebra isomorphism. If, however, we fix a 
lapse $N$ and $k^a,\bar k^a\in{\cal G}_N$, then by (2.1.15) and the 
product law (3.2.8) we have 

$$
\Bigl\{C\bigl[k^a\bigr],C\bigl[\bar k^a\bigr]\Bigr\}=-C\Bigl[\bigl[k,
\bar k\bigr]^a\Bigr]+C\Bigl[\bigl(t^at^b+2q^{ab}\bigr)\bigl(\nu\nabla
_{(b}\bar k_{c)}-\bar\nu\nabla_{(b}k_{c)}\bigr)t^c\Bigr]. \eqno(3.2.9)
$$
\ni
In general the second term on the right is non-zero even for 
asymptotic Killing vectors $k^a,\bar k^a\in{\cal G}^K_\xi$ for some 
$\xi^a$. On the other hand, if we restrict the vector fields $k^a$ 
and $\bar k^a$ further to be in ${\cal G}^0_\xi$ too, then the second 
term in (3.2.9) is vanishing. Thus although ${\cal G}^0_\xi$ is not 
closed respect to the Lie bracket, the restriction of the constraint 
function $C$ to ${\cal G}^0_\xi$ mimics the injective Lie algebra 
(anti-)homomorphisms: the Poisson bracket of $C[k^a]$ and $C[\bar k
^a]$ for $k^a,\bar k^a\in{\cal G}^0_\xi$ is just (minus) the 
constraint function at $[k,\bar k]^a\in{\cal G}_N$. 

To discuss the linear isomorphism $C:{\cal G}\rightarrow{\cal C}$ 
further, recall that the flow on the phase space generated by the 
Hamiltonian vector field $X_{C[k^e]}=(\delta C[k^e]/\delta\tilde p^{ab},
-\delta C[k^e]/\delta q_{ab})$ is the congruence of the integral curves 
of the differential equations 

$$
{{\rm d}q_{ab}(u)\over{\rm d}u}={\delta C[k^e]\over\delta\tilde 
  p^{ab}}, \hskip 20pt 
{{\rm d}\tilde p^{ab}(u)\over{\rm d}u}=-{\delta C[k^e]\over\delta q
_{ab}}. \eqno(3.2.10)
$$
\ni
By (3.2.2), (3.2.3) and (3.1.3), on the constraint surface these are 
precisely the vacuum evolution equations (2.1.2) and (2.1.7) with 
lapse $\nu$ and shift $\nu^a$ with respect to the coordinate time 
$t=u$. Let $\phi_u$ be the local 1-parameter family of diffeomorphisms 
(on an open neighbourhood of $\Sigma$ in the spacetime) generated by 
$k^a\in{\cal G}$. Then, for small enough $u$, its action on the 
canonical variables as fields on this neighbourhood is $q_{ab}\mapsto 
q_{ab}-u\dot q_{ab}$ and $\tilde p^{ab}\mapsto\tilde p^{ab}-u\dot
{\tilde p}{}^{ab}$. Thus the canonical variables $q_{ab}$ and $\tilde 
p^{ab}$ are changing along the flow generated by the Hamiltonian 
vector field of $C[k^a]$ on $\Gamma$ exactly in the same way as under 
the action of the diffeomorphism generated by $-k^a$ on the 
spacetime. Thus one may say that the flow of $X_{C[k^a]}$ on $\Gamma$ 
is the natural lift of the flow of $-k^a$ from the spacetime to the 
constraint surface. In the next subsection we show that the theory's 
gauge transformations on the constraint surface are generated 
precisely by the elements of ${\cal G}$. 

\bigskip

\ni
{\bf 3.3 The gauge transformations}\par
\medskip
\ni
Since the Hamiltonian vector fields $X_{C[k^a]}$, $k^a\in{\cal G}$, 
are tangent to $\Gamma$ on $\Gamma$, they belong to the kernel 
distribution 

$$
{\rm ker}\,\Omega\vert_\Gamma:=\bigl\{(\delta q_{ab},\delta\tilde p
^{ab})\in T\Gamma\,\vert \,\Omega((\delta q_{ab},\delta\tilde p^{ab}),
(\delta'q_{ab},\delta'\tilde p^{ab}))=0\,\,\,\,\forall\,\,\,(\delta'q
_{ab},\delta'\tilde p^{ab})\in T\Gamma\,\bigr\}\eqno(3.3.1)
$$
\ni
of the pull back to $\Gamma$ of the canonical symplectic 2-form 
$\Omega$. Since $\Omega$ is closed, this kernel distribution is 
always integrable. But, by definition, the reduced phase space, 
representing the physical degrees of freedoms, is the pair $(\hat
\Gamma,\hat\Omega)$, where $\hat\Gamma$ is the set of the integral 
submanifolds of ${\rm ker}\,\Omega\vert_\Gamma$ and $\hat\Omega$ is 
the (necessarily well defined) projection of $\Omega$ to $\hat
\Gamma$. Thus any integral submanifold of ${\rm ker}\,\Omega\vert
_\Gamma$ through a given point $(q_{ab},\tilde p^{ab})\in\Gamma$ is 
projected to a single point of $\hat\Gamma$, and hence the vector 
fields on $\Gamma$ belonging to the kernel distribution should be 
interpreted as infinitesimal gauge motions, i.e. generators of gauge 
transformations. Therefore, the Hamiltonian vector fields $X_{C[k^a]}$ 
generate gauge transformations on the constraint surface for any $k
^a\in{\cal G}$. In this subsection we show that the converse of this 
statement is also true, namely that any vector field on $\Gamma$ 
belonging to ${\rm ker}\,\Omega\vert_\Gamma$ (and represented by 
smooth fields on $\Sigma$), i.e. any infinitesimal gauge motion, is 
necessarily a Hamiltonian vector field $X_{C[k^a]}$ for some $k^a\in
{\cal G}$. Thus first let us discuss this kernel. 

By the definition of $\Omega$ the kernel of $\Omega\vert_\Gamma$ at 
$(q_{ab},\tilde p^{ab})$ consists of all the vectors $(\delta q_{ab},
\delta\tilde p^{ab})$ tangent to $\Gamma$ for which $\int_\Sigma(
\delta\tilde p^{ab}\delta'q_{ab}-\delta'\tilde p^{ab}\delta q_{ab})
{\rm d}^nx=0$ for any vector $(\delta'q_{ab},\delta'\tilde p^{ab})$ 
tangent to $\Gamma$ at $(q_{ab},\tilde p^{ab})$. To evaluate this 
condition for $(\delta q_{ab},\delta\tilde p^{ab})$, we should take 
into account that $\delta'q_{ab}$ and $\delta'\tilde p^{ab}$ are not 
independent. In fact, since $\Gamma=\{(q_{ab},\tilde p^{ab})\vert \,
\,\tilde{\cal C}(q_{ef},\tilde p^{ef})=0,\,\,\tilde{\cal C}_a(q_{ef},
\tilde p^{ef})=0\}$, the tangents $(\delta'q_{ab},\delta'\tilde p
^{ab})$ of $\Gamma$ at $(q_{ab},\tilde p^{ab})$ must satisfy the 
`linearized constraint equations' $\delta'\tilde{\cal C}:=({{\rm d}
\over{\rm d}u}\tilde{\cal C}(q_{ef}(u),\tilde p^{ef}(u)))_{u=0}=0$ 
and $\delta'\tilde{\cal C}_a:=({{\rm d}\over{\rm d}u}\tilde{\cal C}
_a(q_{ef}(u),\tilde p^{ef}(u)))_{u=0}=0$. Multiplying them by an 
arbitrary function $\lambda$ and spatial vector field $\lambda^a$, 
respectively, and adding them together we obtain 

$$\eqalign{
0=\lambda\delta'\tilde{\cal C}+\lambda^a\delta'\tilde{\cal C}_a&=
 {\delta C[\lambda,\lambda^e]\over\delta q_{ab}}\delta'q_{ab}+
 {\delta C[\lambda,\lambda^e]\over\delta\tilde p^{ab}}\delta'\tilde 
 p^{ab}+D_e\Bigl\{{1\over2\kappa}\sqrt{\vert q\vert}\delta'q_{ab}
 \bigl(q^{ab}D^e\lambda-q^{ea}D^b\lambda\bigr)-\cr
&-{1\over2\kappa}\sqrt{\vert q\vert}\lambda\bigl(q^{ab}D^e\delta'q
 _{ab}-q^{ea}D^b\delta'q_{ab}\bigr)-2\tilde p^{ea}\lambda^b\delta'q
 _{ab}+\lambda^e\tilde p^{ab}\delta'q_{ab}-2\lambda_a\delta'\tilde p
 ^{ea}\Bigr\},\cr}\eqno(3.3.2)
$$
\ni
where $\delta C[\lambda,\lambda^e]/\delta q_{ab}$ and $\delta C[
\lambda,\lambda^e]/\delta\tilde p^{ab}$ are given formally by (3.2.2) 
and (3.2.3). Adding the integral of (3.3.2) to $\int_\Sigma(\delta
\tilde p^{ab}\delta'q_{ab}-\delta'\tilde p^{ab}\delta q_{ab}){\rm d}
^nx=0$ we obtain 

$$
\int_\Sigma\Bigl\{\Bigl(\delta\tilde p^{ab}+{\delta C[\lambda,\lambda
^e]\over\delta q_{ab}}\Bigr)\delta'q_{ab}-\Bigl(\delta q_{ab}-{\delta 
C[\lambda,\lambda^e]\over\delta\tilde p^{ab}}\Bigr)\delta'\tilde p
^{ab}+D_a\tilde W^a(\lambda,\lambda^e)\Bigr\}{\rm d}^nx=0, \eqno(3.3.3)
$$
\ni
where $D_a\tilde W^a(\lambda,\lambda^e)$ denotes the total divergence 
in (3.3.2). But from (3.3.3) we can read off the vanishing of the 
coefficients of $\delta'q_{ab}$ and $\delta'\tilde p^{ab}$ (by the 
Lagrange lemma of the elementary calculus of variations) only if the 
integral of $D_a\tilde W^a(\lambda,\lambda^e)$ is vanishing. From 
(3.3.3) it follows that it is vanishing if $\lambda$, $\lambda^a$ have 
the asymptotic form (3.1.11), and hence, in particular, $C[\lambda,
\lambda^a]$ exists. Therefore, {\it the generators $\lambda$, $\lambda
^a$ of the infinitesimal gauge transformations are precisely the 
smearing fields}. 

Since the vanishing of the Hamiltonian vector field $X_{C[k^a]}$ 
implies the vanishing of the vector field $k^a$ itself, the gauge 
transformations generated by $C[k^a]$ are effective. The condition 
ensuring that the infinitesimal gauge transformations close to a 
Lie algebra are the stronger fall-off (3.2.7), which are precisely 
the fall-off properties of the elements of ${\cal G}$. 

\bigskip

\ni
{\bf 3.4 The Hamiltonian}\par
\medskip
\ni
The aim of this subsection is a concise rederivation of the 
Hamiltonian of Beig and \'O Murchadha, to determine the exact and 
most general boundary conditions for $M$ and $M^a$ for which the 
Hamiltonian is well defined and differentiable, and to see that $M$ 
and $M^a$ may still be arbitrary functions of time. Thus let us start 
with the `basic Hamiltonian' 

$$
H_0[M,M^a]:=C[M,M^a]=\int_\Sigma\bigl(\tilde{\cal C}M+\tilde{\cal 
C}_aM^a\bigr){\rm d}^nx \eqno(3.4.1)
$$
and determine that total divergence $D_a\tilde Z^a$ for which the 
`total Hamiltonian' $H[M,M^a]:=C[M,M^a]+\int_\Sigma D_a\tilde Z^a{\rm 
d}^nx$ is well defined even for the pairs $(M,M^a)$ defining general 
allowed time axes, or at least asymptotic Killing vectors. Here 
$\tilde Z^a$ is expected to be a local expression of the canonical 
variables, the fields $M$ and $M^a$ and their spatial derivatives up 
to some finite order. 

Let the difference of the physical and background connections be 
characterized by $\Gamma^c_{ab}X^b:=(D_a-{}_0D_a)X^c$, which $\Gamma^c
_{ab}$ is given explicitly by $\Gamma^c_{ab}={1\over2}q^{cd}(-{}_0D_d
q_{ab}+{}_0D_aq_{bd}+{}_0D_bq_{da})$. Then the physical curvature 
scalar of $q_{ab}$ can be written as $R=q^{ab}q^{cd}({}_0D_a\,{}_0D_b
q_{cd}-{}_0D_a\,{}_0D_cq_{bd})+q_{cd}q^{ab}q^{ef}(\Gamma^c_{ab}\Gamma
^d_{ef}-\Gamma^c_{ae}\Gamma^d_{bf})$, and, following Beig and \'O 
Murchadha, we write 

$$\eqalign{
H_0[M,0]=-{1\over2\kappa}\int_\Sigma\Bigl(&Mq^{ab}q^{cd}{}_0D_a\bigl(
 {}_0D_bq_{cd}-{}_0D_cq_{bd}\bigr)+\cr
+&Mq_{cd}q^{ab}q^{ef}\bigl(\Gamma^c_{ab}\Gamma^d_{ef}-\Gamma^c_{ae}
 \Gamma^d_{bf}\bigr)+{4\kappa^2\over\vert q\vert}M\bigl({1\over(n-1)}
 (\tilde p^{ab}q_{ab})^2-\tilde p^{ab}\tilde p_{ab}\bigr)\Bigr)\sqrt{
 \vert q\vert}{\rm d}^nx.\cr}\eqno(3.4.2)
$$
\ni
The integral of the second and third terms on the right is finite if 

$$
M(t,x^{\bk})=2x^{\bk}B_{\bk}(t)+T(t)+r^K\mu^{(K)}(t,{x^{\bk}\over r})
+o^\infty(r^K), \eqno(3.4.3a)
$$
\ni
where $K\leq2k+2-n$, and if the equality holds here then $\mu^{(K)}(t,
{x^{\bk}\over r})$ must be an odd parity function of ${x^{\bk}\over 
r}$. Here we also had to use $k\geq{1\over2}(n-1)$ if $B_{\bi}\not=0$. 
On the other hand, without additional restrictions on $k$ and $n$, 
the first term has finite integral only for those functions 
(3.4.3a) in which both $B_{\bi}$ and $T$ are vanishing. Since the 
first term is not a pure total divergence, we should write this as 
the sum of a total divergence and terms that already yield finite 
integral even for $M$ above. Beig and \'O Murchadha wrote this 
`wrong' term as 

$$\eqalign{
&Mq^{ab}q^{cd}{}_0D_a\bigl({}_0D_bq_{cd}-{}_0D_cq_{bd}\bigr)\sqrt{
 \vert q\vert}=\cr
=&{}_0D_a\Bigl(Mq^{ab}q^{cd}\bigl({}_0D_bq_{cd}-{}_0D_cq_{bd}\bigr)
 \sqrt{\vert q\vert}\Bigr)-M{}_0D_a\bigl(q^{ab}q^{cd}\sqrt{\vert q
 \vert}\bigr)\bigl({}_0D_bq_{cd}-{}_0D_cq_{bd}\bigr)-\cr
&-\bigl({}_0D_aM\bigr)q^{ab}q^{cd}\Bigl({}_0D_b\bigl(q_{cd}-{}_0q_{cd}
 \bigr)-{}_0D_c\bigl(q_{bd}-{}_0q_{bd}\bigr)\Bigr)\sqrt{\vert q\vert}=
 \cr
=&{}_0D_a\Bigl(Mq^{ab}q^{cd}\bigl({}_0D_bq_{cd}-{}_0D_cq_{bd}\bigr)
 \sqrt{\vert q\vert}\Bigr)-M{}_0D_a\bigl(q^{ab}q^{cd}\sqrt{\vert q
 \vert}\bigr)\bigl({}_0D_bq_{cd}-{}_0D_cq_{bd}\bigr)-\cr
&-{}_0D_a\Bigl(\bigl({}_0D_bM\bigr)q^{ab}q^{cd}\bigl(q_{cd}-{}_0q_{cd}
 \bigr)\sqrt{\vert q\vert}-\bigl({}_0D_bM\bigr)q^{bc}q^{ad}\bigl(q
 _{cd}-{}_0q_{cd}\bigr)\sqrt{\vert q\vert}\Bigr)+\cr
&+{}_0D_b\Bigl(\bigl({}_0D_aM\bigr)q^{ab}q^{cd}\sqrt{\vert q\vert}
 \Bigr)\bigl(q_{cd}-{}_0q_{cd}\bigr)-{}_0D_c\Bigl(\bigl({}_0D_aM\bigr)
 q^{ab}q^{cd}\sqrt{\vert q\vert}\Bigr)\bigl(q_{bd}-{}_0q_{bd}\bigr)=
 \cr
=&-D_a\Bigl(Mq^{ab}q^{cd}\bigl({}_0D_cq_{bd}-{}_0D_bq_{cd}\bigr)+
 \bigl({}_0D_bM\bigr)q^{ab}q^{cd}\bigl(q_{cd}-{}_0q_{cd}\bigr)-\bigl(
 {}_0D_cM\bigr)q^{ab}q^{cd}\bigl(q_{bd}-{}_0q_{bd}\bigr)\Bigr)\sqrt{
 \vert q\vert}+\cr
&+{}_0D_a\,{}_0D_bM\Bigl(q^{ab}q^{cd}\bigl(q_{cd}-{}_0q_{cd}\bigr)-
 q^{bd}q^{ac}\bigl(q_{cd}-{}_0q_{cd}\bigr)\Bigr)\sqrt{\vert q\vert}+
 \cr
&+{}_0D_a\bigl(q^{ab}q^{cd}\sqrt{\vert q\vert}\bigr)\Bigl(-M\bigl(
 {}_0D_bq_{cd}-{}_0D_cq_{bd}\bigr)+\bigl({}_0D_bM\bigr)\bigl(q_{cd}-
 {}_0q_{cd}\bigr)-\bigl({}_0D_cM\bigr)\bigl(q_{bd}-{}_0q_{bd}\bigr)
 \Bigr).\cr}\eqno(3.4.4)
$$
\ni
The integral of the second and third terms on the right of (3.4.4) is 
finite if $M$ has the form (3.4.3a) where $K$ satisfies the stronger 
condition $K\leq k+2-n$, and if the equality holds in this inequality 
then $\mu^{(K)}(t,{x^{\bk}\over r})$ is an odd parity function of 
${x^{\bk}\over r}$. Then, since $K\leq k+2-n<2k+2-n$, 

$$\eqalign{
H[M,0]:=C[M,0]-{1\over2\kappa}\int_\Sigma D_a&\Bigl\{Mq^{ab}q^{cd}
 \bigl({}_0D_cq_{bd}-{}_0D_bq_{cd}\bigr)+\bigl({}_0D_bM\bigr)q^{ab}
 q^{cd}\bigl(q_{cd}-{}_0q_{cd}\bigr)-\cr
&-\bigl({}_0D_cM\bigr)q^{ab}q^{cd}\bigl(q_{bd}-{}_0q_{bd}\bigr)\Bigr\}
 {\rm d}\Sigma \cr}\eqno(3.4.5)
$$
\ni
is already well defined. If $(\delta q_{ab},\delta\tilde p^{ab})$ is 
any tangent vector at $(q_{ab},\tilde p^{ab})\in T^*{\cal Q}$, then 
the derivative of $H[M,0]$ in the direction $(\delta q_{ab},\delta
\tilde p^{ab})$ is 

$$\eqalign{
\delta H[M,0]=&\int_\Sigma\Bigl({\delta C[M,0]\over\delta q_{ab}}
 \delta q_{ab}+{\delta C[M,0]\over\delta\tilde p^{ab}}\delta\tilde 
 p^{ab}\Bigr){\rm d}^nx-{1\over2\kappa}\int_\Sigma D_a\Bigl\{Mq^{ab}
 q^{cd}\sqrt{\vert q\vert}\bigl(\Gamma^e_{dc}\delta q_{be}-\Gamma^e
 _{db}\delta q_{ce}\bigr)+\cr
&+\delta\bigl(q^{ab}q^{cd}\sqrt{\vert q\vert}\bigr)\Bigl(M\bigl({}_0D
 _cq_{bd}-{}_0D_bq_{cd}\bigr)+\bigl({}_0D_bM\bigr)\bigl(q_{cd}-{}_0q
 _{cd}\bigr)-\bigl({}_0D_cM\bigr)\bigl(q_{bd}-{}_0q_{bd}\bigr)\Bigr)
\Bigr\}{\rm d}^nx,\cr}\eqno(3.4.6)
$$
\ni
where $\delta C[M,0]/\delta q_{ab}$ and $\delta C[M,0]/\delta\tilde 
p^{ab}$ are given formally by (3.2.2) and (3.2.3). (Strictly speaking, 
these are {\it not} functional derivatives of $C[M,0]$, because the 
constraint function $C$ is not well defined for $M$ above.) Since, 
however, the integral of the total divergence on the right of (3.4.6) 
is vanishing for the functions $M$ given by (3.4.3a), $H[M,0]$ is 
functionally differentiable with respect to the canonical variables 
too. 

We can write 

$$
H_0[0,M^a]=-2\int_\Sigma\bigl(D_a\tilde p^{ab}\bigr)M_b{\rm d}^nx=
2\int_\Sigma\Bigl(\tilde p^{ab}{}_0D_{(a}M_{b)}-\tilde p^{ab}\Gamma
^c_{ab}M_c-D_a\bigl(\tilde p^{ab}M_b\bigr)\Bigr){\rm d}^nx. 
\eqno(3.4.7)
$$
The integral of the first two terms on the right is well defined even 
for vector fields $M^a$ of the form 

$$
M_{\bi}(t,x^{\bk})=2x^{\bk}R_{\bk\bi}(t)+T_{\bi}(t)+r^L\mu^{(L)}
_{\bi}(t,{x^{\bk}\over r})+o^\infty(r^L), \eqno(3.4.3b)
$$
\ni
where $L\leq k+2-n$, and if the equality holds in this inequality then 
$\mu^{(L)}_{\bi}(t,{x^{\bk}\over r})$ is an odd parity function of 
${x^{\bk}\over r}$. Note that, to prove the existence of the integral 
of the first two terms for nonzero $R_{\bi\bj}$ in (3.4.3b) we also 
had to use $k\geq{1\over2}(n-1)$. The integral of the third term on 
the right of (3.4.7) is, however, finite only for $R_{\bk\bi}(t)=0$ 
and $T_{\bi}(t)=0$. Thus 

$$
H[0,M^a]:=C[0,M^a]+2\int_\Sigma D_a\bigl(\tilde p^{ab}M_b\bigr)
{\rm d}^nx=\int_\Sigma \tilde p^{ab}\L_{\bf M}q_{ab}\,{\rm d}^nx
\eqno(3.4.8) 
$$
\ni
is well defined even for vector fields $M^a$ with the structure 
(3.4.3b). The derivative of $H[0,M^a]$ in the direction $(\delta q
_{ab},\delta\tilde p^{ab})$ is 

$$
\delta H[0,M^a]=\int_\Sigma\Bigl(\L_{\bf M}q_{ab}\,\delta\tilde p
^{ab}-\L_{\bf M}\tilde p^{ab}\,\delta q_{ab}+D_e\bigl(M^e\tilde 
p^{ab}\delta q_{ab}\bigr)\Bigr){\rm d}^nx. \eqno(3.4.9)
$$
\ni
Since for the vector fields $M^a$ above the integral of the total 
divergence is zero, $H[0,M^a]$ is differentiable with respect to the 
canonical variables too. 

Therefore, the Hamiltonian $H[M,M^a]:=H[M,0]+H[0,M^a]$ of Beig and 
\'O Murchadha is finite and functionally differentiable with respect 
to the canonical variables even for $M$ and $M^a$ given by (3.4.3) 
with $K,L\leq k+2-n$, and if the equality holds in these inequalities 
then $\mu^{(K)}$ and $\mu^{(L)}_{\bi}$ must be odd parity functions 
of ${x^{\bk}\over r}$, respectively. However, the spacetime vector 
field $K^a:=Mt^a+M^a$ is still {\it not} needed to be an asymptotic 
Killing vector field with respect to some foliation (and not even an 
allowed time axis), because the fall-off rates $K$ and $L$ are still 
required only to satisfy $K,L\leq k+2-n$ (instead of $K,L\leq(1-k)$). 
Moreover, $R_{\bi\bj}$, $B_{\bi}$, $T_{\bi}$ and $T$ may still have 
arbitrary time dependence. 

\bigskip

\ni
{\bf 3.5 The algebra of the Hamiltonians and the asymptotic 
symmetries}\par
\medskip
\ni
On $\Gamma$ the system of equations 

$$
{{\rm d}q_{ab}\over{\rm d}t}={\delta H[M,M^e]\over\delta\tilde p
^{ab}},\hskip 20pt
{{\rm d}\tilde p^{ab}\over{\rm d}t}=-{\delta H[M,M^e]\over\delta q
_{ab}},\eqno(3.5.1)
$$
\ni
defining the Hamiltonian flow on $\Gamma$, is precisely the system 
of the vacuum evolution equations with lapse $M$ and shift $M^a$. 
However, this system still does {\it not} preserve the boundary 
conditions (2.3.1a-3a) and (3.1.4-6) for the canonical variables, 
because the regularity and functional differentiability of the 
Hamiltonian $H[M,M^a]$ implied only $K,L\leq k+2-n$, which is weaker 
than $K,L\leq(1-k)$. Thus, based on the analysis of subsection 2.3, 
{\it we must require that the powers $K$ and $L$ satisfy} 

$$
K,L\leq(1-k). \eqno(3.5.2)
$$
\ni
Therefore, $K^a=Mt^a+M^a$ already has the asymptotic form (2.4.1), 
i.e. $K^a$ must be an element of ${\cal A}$, and the restriction 
of the Hamiltonian $H$ to the pure gauge generators $\mu$ and $\mu^a$ 
(i.e. for which $\mu t^a+\mu^a\in{\cal G}$), is just $C[\mu,\mu^a]$. 

Repeating the analysis of subsection 3.2, one can show that (3.5.2) 
ensures the vanishing of the boundary terms appearing in the 
calculation of the Poisson brackets of two Hamiltonians, $H[M,M^a]$ 
and $H[\bar M,\bar M^a]$, and, by (2.4.8-10), that $M^aD_a\bar M-\bar 
M^aD_aM$ has the structure of a lapse and $[M,\bar M]^a$ and $\bar M
D^aM-MD^a\bar M$ have the structure of a shift satisfying (3.5.2). 
The resulting Poisson algebra of the Hamiltonians [5] (see also 
[4,28]) is 

$$
\Bigl\{H\bigl[M,M^a\bigr],H\bigl[\bar M,\bar M^a\bigr]\Bigr\}=-H\Bigl[
\L_{\bf M}\bar M-\L_{\bar{\bf M}}M,\bigl[M,\bar M\bigr]^a-\bigl(MD^a
\bar M-\bar MD^aM\bigr)\Bigr].\eqno(3.5.3)
$$
\ni
For pure gauge generators (3.5.3) reduces to (3.2.8), and the 
Poisson bracket of $H[M,M^a]$ and $C[\nu,\nu^a]$ is also a constraint 
function. Thus the Poisson algebra ${\cal C}$ of the constraint 
functions is an ideal in the Poisson algebra ${\cal H}$ of the 
Hamiltonians parameterized by the allowed time axes $K^a\in{\cal A}$. 
The structure of ${\cal H}$ can be determined easily by considering 
the Hamiltonians parameterized by the special allowed time axes like 
$K^a=2x^{\bk}B_{\bk}(t)t^a$, $K^a=2x^{\bj}R_{\bj}{}^{\bi}(t)({\partial
\over\partial x^{\bi}})^a$, ... etc. With this parameterization of the 
Hamiltonians (3.5.3) shows that, for {\it each fixed value $t$} of 
the coordinate time, the factor of ${\cal H}$ with the ideal ${\cal 
C}$ is just the Poincare algebra. Thus ${\cal H}/{\cal C}$ is infinite 
dimensional. If, however, the coefficients $B_{\bi}$, $R_{\bi\bj}$, 
$T_{\bi}$ and $T$ are restricted to be the coefficients in the 
asymptotic Killing vectors with respect to some $\xi^a$ as in 
subsection 2.4, then the whole factor ${\cal H}/{\cal C}$ would be 
finite dimensional, and, in fact, the Poincare algebra. 

As we noted in subsection 2.4, the space ${\cal A}$ of the allowed 
time axes does not form a Lie algebra with the natural Lie bracket in 
general. Hence the Poisson algebra ${\cal H}$ of all the Hamiltonians, 
indexed by the elements of ${\cal A}$, does not seem to be connected 
in a natural way to some naturally defined Lie algebra of spacetime 
vector fields. However, restricting the spacetime vector fields $K^a$ 
and $\bar K^a$ to be from the subspace ${\cal A}^0_\xi$ of the space 
${\cal A}^K_\xi$ of the asymptotic Killing vectors for some $\xi^a$ 
and writing $H[K^a]:=H[M,M^a]$, by (2.1.15) the product law (3.5.3) 
takes the remarkably simple form 

$$
\Bigl\{H\bigl[K^a\bigr],H\bigl[\bar K^a\bigr]\Bigr\}=-H\Bigl[\bigl[K,
\bar K\bigr]^a\Bigr].\eqno(3.5.4)
$$
\ni
If $K^a$ and $\bar K^a$ are allowed to be from ${\cal A}^K_\xi$, 
then the first two terms on the right of (2.1.15) give a constraint 
function with uncontrollable generators, as in (3.2.9), and we have 
only 

$$
\Bigl\{H\bigl[K^a\bigr],H\bigl[\bar K^a\bigr]\Bigr\}+H\Bigl[\bigl[K,
\bar K\bigr]^a\Bigr]\in C\bigl({\cal G}\bigr)={\cal C}. \eqno(3.5.5)
$$
\ni
Therefore, {\it the set ${\cal H}_\xi:=H({\cal A}^K_\xi)$ of the 
Hamiltonians parameterized by the asymptotic Killing vectors $K^a$ 
with respect to $\xi^a$ is the Lie product preserving image of 
${\cal A}^K_\xi$ modulo constraints, and on the elements of the 
subspace ${\cal A}^0_\xi$ the Hamiltonian $H$ preserves the Lie 
product}. 

Finally, let $M$, $N$, $N^a$ and $M^a$ be as in (2.4.1), and calculate 
the {\it total} time derivative of $H[M,M^a]$ along $\xi^a:=Nt^a+N^a$. 
Since $M$ and $M^a$ may depend on $t$, the derivative consists of two 
terms: 

$$\eqalign{
{{\rm d}\over{\rm d}t}H\bigl[M,M^e\bigr]&=H\bigl[\dot M,\dot M^e\bigr]
 +\int_\Sigma\Bigl({\delta H[M,M^e]\over\delta\tilde p^{ab}}\dot
 {\tilde p}{}^{ab}+{\delta H[M,M^e]\over\delta q_{ab}}\dot q_{ab}
 \Bigr){\rm d}^nx=\cr
&=H\bigl[\dot M,\dot M^e\bigr]+\Bigl\{\,H\bigl[N,N^e\bigr],H\bigl[M,
 M^e\bigr]\,\Bigr\}=\cr
&=H\bigl[\dot M+\L_{\bf M}N-\L_{\bf N}M,\dot M^e+ND^eM-MD^eN-[N,M]^e
 \bigr].\cr}\eqno(3.5.6)
$$
\ni
Here first we used (3.5.1) (which, on the constraint surface, are 
the vacuum evolution equations), and then (3.5.3). If $K^a\in{\cal A}
^0_\xi$, then by (2.1.13) and (2.1.14) the right hand side is zero, 
while for $K^a\in{\cal A}^K_\xi$ the right hand side is a constraint 
function by the asymptotic Killing equations (2.4.2) and (2.4.3). 
Therefore, {\it the Hamiltonian of Beig and \'O Murchadha is constant 
along any allowed time axis $\xi^a$ modulo constraints for the 
asymptotic Killing vectors $K^a\in{\cal A}^K_\xi$, and it is strictly 
constant for vectors $K^a$ that are strong asymptotic Killing with 
respect to the time axis $\xi^a$}. 

To summarize: First, to ensure that e.g. the symplectic 2-form be well 
defined, in addition to the result $l=k+1$ of the analysis of 
subsection 2.3, we had to assume that $k\geq{1\over2}(n-1)$. Then 
the constraint functions are well defined, functionally differentiable 
and close to a Lie algebra precisely for those smearing fields that 
correspond to the elements of ${\cal G}$ itself. The constraint 
function preserves the Lie product of the elements of the space ${\cal 
G}^0_\xi$ in the Poisson algebra. 
The Hamiltonian of Beig and \'O Murchadha, parameterized by the allowed 
time axes $K^a$, are finite, functionally differentiable, close to an 
infinite dimensional Lie algebra, and the Hamilton equations preserve 
the boundary conditions for the canonical variables. 
The Hamiltonian preserves the Lie product of the elements of the space 
${\cal A}^0_\xi$, and preserves the Lie product of the elements of the 
space ${\cal A}^K_\xi$ modulo constraints. 
The Beig--\'O Murchadha Hamiltonian $H[K^a]$ is constant in time with 
respect to $\xi^a$ for any $K^a\in{\cal A}^0_\xi$, but it is only 
constant modulo constraints for the asymptotic Killing vectors $K^a
\in{\cal A}^K_\xi$. 

\bigskip
\bigskip
\ni
{\lbf 4. The ADM conserved quantities of matter+gravity systems}\par
\medskip
\ni
{\bf 4.1 The ADM conserved quantities}\par
\medskip
\ni
In the complete Hamiltonian description the matter fields would have 
to be included. However, a detailed Hamiltonian analysis of the 
matter fields is not needed if we are interested only in the ADM 
energy-momentum and angular momentum, because the value of the 
Hamiltonian of the matter fields on the matter constraint surface 
(if there is any, as in the spacial case of Yang--Mills fields) is 
expected to be $Q_m[K^a]$ for some allowed time axis $K^a\in{\cal A}$. 
Then the gravitational constraint, i.e. a part of Einstein's equation, 
is $Q_m[K^a]+C[K^a]=0$. Thus, the value of the total Hamiltonian on 
the constraint surface, $H[K^a]\vert_{\Gamma}+Q_m[K^a]$, is given by 
the same surface term as in the vacuum case, and it has the structure 
$H[K^a]\vert_{\Gamma}+Q_m[K^a]=T(t)p^0+T_{\bi}(t)p^{\bi}+R_{\bi\bj}(t)
j^{\bi\bj}+2B_{\bi}(t)j^{\bi 0}$, where $T(t)$, $T_{\bi}(t)$, $R_{\bi
\bj}(t)$ and $B_{\bi}(t)$ are the functions appearing e.g. in the form 
(2.4.1) of the allowed time axis $K^a$. If, however, $K^a\in{\cal A}^K
_\xi$ for some $\xi^a\in{\cal A}$, whenever the functions $T(t)$, $T
_{\bi}(t)$, $R_{\bi\bj}(t)$ and $B_{\bi}(t)$ can be represented by the 
${1\over2}m(m+1)$ {\it independent parameters} $T$, $T_{\bi}$, $R_{\bi
\bj}$ and $B_{\bi}$, the evaluation of $H[K^a]+Q_m[K^a]$ on $\Gamma$ 
above defines a linear mapping ${\cal A}^K_\xi/{\cal G}^K_\xi\approx
{\cal A}^0_\xi/{\cal G}^0_\xi\rightarrow{\bf R}$, whose components 
${\tt P}^0$, ${\tt P}^{\bi}$, ${\tt J}^{\bi\bj}$ and ${\tt J}^{\bi 0}$ 
define the energy, the linear momentum, the spatial angular momentum 
and centre-of-mass, respectively, via $H[K^a]\vert_{\Gamma}+Q_m[K^a]=:
T{\tt P}^0+T_{\bi}{\tt P}^{\bi}+R_{\bi\bj}{\tt J}^{\bi\bj}+2B_{\bi}
{\tt J}^{\bi 0}$. 
In particular, if $K^a\in{\cal A}^K_\xi$ and $\xi^a\in{\cal G}$, 
whenever the functions $T(t)$, $T_{\bi}(t)$, $R_{\bi\bj}(t)$ and $B
_{\bi}(t)$ are all constant, then, as we will see, ${\tt P}^0$ and 
${\tt P}^{\bi}$ are just the familiar ADM energy and linear momentum, 
respectively, ${\tt J}^{\bi\bj}$ is the angular momentum of Regge and 
Teitelboim and ${\tt J}^{\bi 0}$ is the centre-of-mass given by Beig 
and \'O Murchadha. 
If, however, $K^a\in{\cal A}^K_\xi$ but $\xi^a-t^a\in{\cal G}$, 
whenever $T(t)$, $R_{\bi\bj}(t)$ and $B_{\bi}(t)$ are still constant 
but $T_{\bi}(t)$ changes in time as $T_{\bi}(t)=T_{\bi}-2tB_{\bi}$ 
(just according to the expression of the $n+1$ form of the boost 
Killing vectors of the Minkowski spacetime), then ${\tt P}^0$, ${\tt 
P}^{\bi}$ and ${\tt J}^{\bi\bj}$ remains the same but the 
centre-of-mass ${\tt J}^{\bi 0}$ deviates from that of Beig and \'O 
Murchadha by the term $t{\tt P}^{\bi}$. 
Since ${\tt P}^{\ua}$ and ${\tt J}^{\ua\ub}$, ${\ua},{\ub}=0,1,...,n$, 
are elements of the dual space of ${\cal A}^K_\xi/{\cal G}^K_\xi$, 
it is easy to check that under a Lorentz transformation $x^{\ua}
\mapsto x^{\ub}\Lambda_{\ub}{}^{\ua}$ of the Cartesian coordinates 
$x^{\ua}=(t,x^{\bi})$ the energy-momentum ${\tt P}^{\ua}$ transforms 
as a Lorentz vector and the angular momentum ${\tt J}^{\ua\ub}$ as 
an anti-symmetric tensor: ${\tt P}^{\ua}\mapsto {\tt P}^{\ub}\Lambda
_{\ub}{}^{\ua}$ and ${\tt J}^{\ua\ub}\mapsto {\tt J}^{\uc\ud}\Lambda
_{\uc}{}^{\ua}\Lambda_{\ud}{}^{\ub}$, while under a translation 
$x^{\ua}\mapsto x^{\ua}+\eta^{\ua}$ of the Cartesian coordinates 
${\tt P}^{\ua}$ remains intact and ${\tt J}^{\ua\ub}\mapsto {\tt J}
^{\ua\ub}+2\eta^{[{\ua}}{\tt P}^{{\ub}]}$. Thus under a Poincare 
transformation of the Cartesian coordinates ${\tt P}^{\ua}$ transforms 
like the energy-momentum vector and ${\tt J}^{\ua\ub}$ as the angular 
momentum tensor of a Poincare invariant system. We emphasize that to 
derive these transformation properties the centre-of-mass expression 
of Beig and \'O Murchadha had to be completed by the term $t{\tt P}
^{\bi}$. It might be worth noting that in the so-called field 
formulation of general relativity on a given background [29,30] the 
centre-of-mass expression also contains the extra term $t{\tt P}^{\bi}$. 

The general expression of these quantities in terms of the metric 
and the extrinsic curvature for any allowed time axis $K^a$ is 

$$\eqalign{
Q[K^a]:=H[K^a]\vert_{\Gamma}+Q_m[K^a]=-{1\over2\kappa}\int_\Sigma
 D_a&\Bigl\{Mq^{ab}q^{cd}\bigl({}_0D_cq_{bd}-{}_0D_bq_{cd}\bigr)-
 \bigl({}_0D_bM\bigr)q^{ab}q^{cd}\bigl(q_{cd}-{}_0q_{cd}\bigr)+\cr
&+\bigl({}_0D_cM\bigr)q^{ab}q^{cd}\bigl(q_{bd}-{}_0q_{bd}\bigr)-2M_b
 \bigl(\chi^{ba}-\chi q^{ba}\bigr)\Bigr\}{\rm d}\Sigma.\cr}\eqno(4.1.1)
$$
\ni
Therefore, we can {\it define} the energy-momentum and angular 
momentum of any asymptotic end by the surface term of (4.1.1) even in 
the presence of matter fields, independently of any symplectic or 
Hamiltonian structure or phase space. The only requirement is its 
existence, and we assume only that the boundary conditions obtained 
from the investigations of the evolution equations in subsection 2.3 
hold. Apparently, for asymptotic boosts (i.e. for $B_{\bi}\not=0$) 
and rotations ($R_{\bi\bj}\not=0$) (4.1.1) gives finite value only if 
$k=n-1$, and for asymptotic translations (i.e. for $B_{\bi}=0$, $R
_{\bi\bj}=0$ but $T\not=0$ or $T_{\bi}\not=0$) only if $k=n-2$. 
However, it is well known that the ADM energy-momentum is finite and 
well defined even if the metric falls off only slightly faster than 
$r^{-{1\over2}(n-2)}$, because the contribution of the slow fall-off 
`part' of the metric and extrinsic curvature to the ADM energy-momentum 
can always be written as a constraint: the whole Hamiltonian expressed 
as a volume integral is finite even for the slower fall-off asymptotic 
ends [10]. We show that, by the same reason, the angular momentum and 
centre-of-mass can be finite even for metrics with $r^{-{1\over2}
(n-1)}$ fall-off. In particular, in $m=3+1$ spacetime dimensions the 
{\it a priori} $1/r$ fall-off of Beig and \'O Murchadha is the weakest 
possible for which, in general, we can have finite angular momentum.

To determine the weakest possible power-type boundary conditions coming 
from the finiteness of $Q[K^a]$, let us rewrite it as an integral on 
$\Sigma$ by the very definition (3.4.5) and (3.4.8) of the Hamiltonian:

$$\eqalign{
Q[K^a]=\int_\Sigma\bigl(&j^aM_a+\mu M\bigr){\rm d}\Sigma
+{1\over\kappa}\int_\Sigma\bigl(\chi^{ab}-\chi q^{ab}\bigr)\bigl({}_0
 D_{(a}M_{b)}-\Gamma^c_{ab}M_c\bigr)\,{\rm d}\Sigma-\cr
-{1\over2\kappa}\int_\Sigma\Bigl\{&{}_0D_a\,{}_0D_bM\Bigl(q^{ab}q^{cd}
 \bigl(q_{cd}-{}_0q_{cd}\bigr)-q^{bd}q^{ac}\bigl(q_{cd}-{}_0q_{cd}
 \bigr)\Bigl)\sqrt{\vert q\vert}+\cr
&+{}_0D_a\bigl(q^{ab}q^{cd}\sqrt{\vert q\vert}\bigr)\Bigl(\bigl({}_0
 D_bM\bigr)\bigl(q_{cd}-{}_0q_{cd}\bigr)-\bigl({}_0D_cM\bigr)\bigl(
 q_{bd}-{}_0q_{bd}\bigr)\Bigr)-\cr
&-M{}_0D_a\bigl(q^{ab}q^{cd}\sqrt{\vert q\vert}\bigr)\bigl({}_0D_bq
 _{cd}-{}_0D_cq_{bd}\bigr)+\cr
&+Mq_{cd}q^{ab}q^{ef}\bigl(\Gamma^c_{ab}\Gamma^d_{ef}-\Gamma^c_{ae}
 \Gamma^d_{bf}\bigr)\sqrt{\vert q\vert}+M\bigl(\chi^2-\chi_{ab}\chi
 ^{ab}\bigr)\sqrt{\vert q\vert}\Bigr\}{\rm d}^nx, \cr}\eqno(4.1.2) 
$$
\ni
where $M$ and $M_a$ have the form (2.4.1) for some (unspecified) 
powers $E$ and $F$. Suppose that the energy density and the momentum 
density of the matter fields satisfy the fall-off and parity 
conditions of subsection 2.3. If $B_{\bi}\not=0$ then the condition of 
the existence of the integrals involving $M$ is $E\leq(1-k)$ and 
$k\geq{1\over2}(n-1)$, and if the equality $E=(1-k)$ holds then $\mu
^{(E)}(t,{x^{\bk}\over r})$ of (2.4.1) has odd parity. Thus, in 
particular, $k\geq1$ and $E\leq0$ must hold if $n\geq3$. If $B_{\bi}
=0$ then the rate $k$ of the fall-off can be reduced. In fact, the 
condition of the existence of the integrals involving $M$ is $E\leq-k$ 
and $k>{1\over2}(n-2)$, which, for $n=3$, gives the well known results 
$k>{1\over2}$ and $E\leq-k<-{1\over2}$. Similarly, if $R_{\bi\bj}
\not=0$ then the condition of the finiteness of the integrals 
involving $M_a$ is $k\geq{1\over2}(n-1)$ and $F\leq(1-k)$, and if the 
equality $F=(1-k)$ holds then $\mu^{(F)}_{\bi}(t,{x^{\bk}\over r})$ 
of (2.4.1) has odd parity. If $R_{\bi\bj}=0$ then the fall-off may be 
slower: $k>{1\over2}(n-2)$ and $F\leq-k$. Therefore, {\it the 
energy-momentum and (relativistic) angular momentum are finite for 
general time axes $K^a$ precisely when $k\geq{1\over2}(n-1)$, but for 
the slower fall-off $k>{1\over2}(n-2)$ the finiteness of the 
energy-momentum is not guaranteed by $B_{\bi}=0$ and $R_{\bi\bj}=0$ 
alone, $E,F\leq-k$ must also be required}. 

This motivates us to consider, for some $q\leq(1-k)$, the special time 
axes $K^a=Mt^a+M^a\in{\cal A}$ with the asymptotic structure 

$$\eqalign{
M&=T(t)+r^E\mu^{(E)}\bigl(t,{x^{\bk}\over r}\bigr)+o^\infty\bigl(r^E
 \bigr), \cr
M_{\bi}&=T_{\bi}(t)+r^F\mu_{\bi}{}^{(F)}\bigl(t,{x^{\bk}\over r}\bigr)
 +o^\infty\bigl(r^F\bigr), \hskip 20pt E,F\leq q. \cr}\eqno(4.1.3)
$$
\ni
A simple calculation shows that they do not form a Lie algebra. If, 
however, they are assumed to be asymptotic Killing vectors too 
(whenever the powers $P$ and $Q$ in the asymptotic Killing equations 
(2.4.2) and (2.4.3) should be required to satisfy $P,Q\leq q-1$, and 
$T_{\bi}$ and $T$ are necessarily constant), then for certain values 
of $q$ they form a subspace in ${\cal A}^K_\xi$ which behaves like 
an ideal of a Lie algebra. In fact, if ${}_q{\cal T}^K_\xi$ is the 
set of the asymptotic Killing vectors $K^a=Mt^a+M^a$ whose components 
satisfy (4.1.3), then by a calculation similar to (2.4.8-10) shows 
that the Lie bracket of $K^a\in{}_q{\cal T}^K_\xi$ and $\bar K^a\in
{\cal A}^K_\xi$ (where the latter is given by (2.4.1)) contains terms 
of order $r^{-k}$. Thus the Lie bracket operation preserves the index 
$q$ of ${}_q{\cal T}^K_\xi$ and the components of $[K,\bar K]^a$ have 
the structure (2.1.3) provided $q\geq-k$. 
The quotient ${}_q{\cal T}^K_\xi/{}_q{\cal T}^K_\xi\cap{\cal G}^K
_\xi$ is isomorphic to ${\bf R}^m$ and inherits a commutative Lie 
algebra structure from ${\cal A}^K_\xi/{\cal G}^K_\xi$. Thus ${}_q
{\cal T}^K_\xi$ may be interpreted as the space of the `$q$ fast 
fall-off' asymptotic translations in ${\cal A}^K_\xi$, where $-k
\leq q\leq(1-k)$. These asymptotic translations can be singled out 
even if $k\in(0,1)$, whenever ${\cal A}^K_\xi/{\cal G}^K_\xi$ is 
only the Lorentz Lie algebra rather than the Poincare one. The 
results of the previous paragraph show that for $k\geq{1\over2}(n-1)$ 
the space of the translations could be any of ${}_q{\cal T}^K_\xi$, 
$q\in[-k,1-k]$, but for $k>{1\over2}(n-2)$ it is just the space ${}
_{-k}{\cal T}^K_\xi$ whose elements yield finite energy-momentum. 

For the sake of logical completeness one should note that, strictly 
speaking, the standard expression for the ADM energy-momentum and 
angular momentum (including the Beig--\'O Murchadha centre-of-mass) 
differs from that given by (4.1.1). However, it is easy to see that 
(4.1.1) coincides with the standard one. In fact, for example the 
first term of the integral (4.1.1) can be written as 

$$\eqalign{
\int_\Sigma D_a&\Bigl\{Mq^{ab}q^{cd}\bigl({}_0D_cq_{bd}-{}_0D_bq_{cd}
 \bigr)\Bigr\}\sqrt{\vert q\vert}{\rm d}^nx=\cr
=\int_\Sigma\,{}_0 D_a&\Bigl\{M\bigl({}_0q^{ab}-r^{-k}q^{(k)ab}+o
 ^\infty\bigl(r^{-k}\bigr)\bigr)\bigl({}_0q^{cd}-r^{-k}q^{(k)cd}+o
 ^\infty\bigl(r^{-k}\bigr)\bigr)\times\cr
&\times\bigl({}_0D_cq_{bd}-{}_0D_bq_{cd}\bigr)\sqrt{1+r^{-k}d+o^\infty
 \bigl(r^{-k}\bigr)}\Bigr\}\sqrt{\vert\,{}_0 q\vert}{\rm d}^nx=\cr
=-\lim_{r\mapsto\infty}\oint_{{\cal S}_r}&\Bigl(M\,{}_0q^{cd}\bigl({}
 _0D_cq_{ad}-\,{}_0D_aq_{cd}\bigr)+r^{-k}F_a\Bigr)\,{}_0v^ar^{n-1}
 {\rm d}{\cal S},
\cr}
$$
\ni
where $q^{(k)ab}:=\,{}_0q^{ac}\,{}_0q^{bd}q^{(k)}_{cd}$, $d$ is a 
smooth function, ${}_0v^a$ is the outward directed ${}_0q
_{ab}$-orthogonal unit normal to the large sphere ${\cal S}_r$ of 
coordinate radius $r$, ${\rm d}{\cal S}$ is the unit sphere volume 
element, and the 1-form $F_a$ is defined by the last equality of 
the integrands. 
If $K^a\in{\cal A}$ and $k\geq{1\over2}(n-1)$, then $F_a\,{}_0v^a=
O(r^{-k})$ holds and its parity is {\it odd}, while if $K^a$ has the 
structure (4.1.3) with $q=-k$ and $k>{1\over2}(n-2)$, then $F_a\,{}
_0v^a=O(r^{-k-1})$. However, in both cases the $r\rightarrow\infty$ 
limit of the integral of the term $r^{n-k-1}F_a\,{}_0v^a$ is zero. 
Similarly, all the remaining terms in (4.1.1) can also be written 
into the form being {\it linear} in the physical metric $q_{ab}$ and 
the extrinsic curvature $\chi_{ab}$, yielding the familiar 
expression given in [5]. 

Finally, calculate the total time derivative of $Q[K^a]$ along any 
allowed time axis $\xi^a$, where $K^a\in{\cal A}_N$. However, in the 
present calculations we cannot use (3.5.1), because they are the 
{\it vacuum} evolution equations in the {\it Hamiltonian phase space}. 
In the presence of the matter in the {\it spacetime} we must use 
(2.1.2) and (2.1.7). They, the definitions and formulae (3.1.3) and 
(3.2.2) imply that 

$$
\dot{\tilde p}{}^{ab}={\delta H[N,N^e]\over\delta q_{ab}}+{1\over2}N
q^{ab}\Bigl({1\over2\kappa}\bigl(R+\chi^2-\chi_{cd}\chi^{cd}\bigr)-\mu
\Bigr)\sqrt{\vert q\vert}-{1\over2}N\sigma^{ab}\sqrt{\vert q\vert}. 
\eqno(4.1.4)
$$
\ni
Then by (2.3.5), (3.5.3), (4.1.4) and the definitions we have 

$$\eqalign{
{{\rm d}\over{\rm d}t}\Bigl(H\bigl[M,M^e\bigr]+&Q_m\bigl[M,M^e\bigr]
 \Bigr)=\cr
=&{{\rm d}\over{\rm d}t}Q_m\bigl[M,M^e\bigr]+H\bigl[\dot M,\dot M^e
 \bigr]+\int_\Sigma\Bigl({\delta H[M,M^e]\over\delta\tilde p^{ab}}
 \dot{\tilde p}{}^{ab}+{\delta H[M,M^e]\over\delta q_{ab}}\dot q_{ab}
 \Bigr){\rm d}^nx=\cr
=&Q_m\bigl[\dot M+\L_{\bf M}N-\L_{\bf N}M,\dot M^e+ND^eM-MD^eN-[N,M]
 ^e\bigr]+H\bigl[\dot M,\dot M^e\bigr]+\cr
&+\Bigl\{\,H\bigl[N,N^e\bigr],H\bigl[M,M^e\bigr]\,\Bigr\}+\int_\Sigma
 \Bigl({1\over2\kappa}\bigl(R+\chi^2-\chi_{ab}\chi^{ab}\bigr)-\mu\Bigr)
 \bigl(M\chi+D_cM^c\bigr)N{\rm d}\Sigma=\cr
=&H\bigl[\dot M+\L_{\bf M}N-\L_{\bf N}M,\dot M^e+ND^eM-MD^eN-[N,M]^e
 \bigr]+\cr
&+Q_m\bigl[\dot M+\L_{\bf M}N-\L_{\bf N}M,\dot M^e+ND^eM-MD^eN-[N,M]^e
 \bigr]+\cr
&+\int_\Sigma\Bigl(
 {1\over2\kappa}\bigl(R+\chi^2-\chi_{ab}\chi^{ab}\bigr)-\mu\Bigr)\bigl(
 M\chi+D_cM^c\bigr)N{\rm d}\Sigma.\cr}
$$
\ni
Taking into account the Lagrangian constraints (2.1.5) and 
(2.1.6) we obtain 

$$
{{\rm d}\over{\rm d}t}Q\bigl[M,M^e\bigr]=Q\bigl[\dot M+\L_{\bf M}N-\L
_{\bf N}M,\dot M^e+ND^eM-MD^eN-[N,M]^e\bigr]. \eqno(4.1.5)
$$
\ni
By (4.1.1) the right hand side is an $(n-1)$-sphere integral at 
infinity with the generators $\tilde M:=\dot M+\L_{\bf M}N-\L_{\bf N}
M$ and $\tilde M^e:=\dot M^e+ND^eM-MD^eN-[N,M]^e$. If $K^a$ is a 
general asymptotic Killing vector, then by (2.4.2) and (2.4.3) the 
order of these generators is at most $O(r^{1-k})$, whenever their 
parity is odd and $k\geq{1\over2}(n-1)$. If $K^a$ is an asymptotic 
translation from ${}_{-k}{\cal T}^K_\xi$ and $k>{1\over2}(n-2)$, 
then the order of these generators is at most $O(r^{-k})$. However, 
in both cases the right hand side of (4.1.5) is vanishing. 
Therefore, {\it the quantities $Q[K^a]$ are constant in time for the 
asymptotic Killing vectors $K^a\in{\cal A}^K_\xi$}. This result is 
analogous to the fact that the $Q_m[K^a]$ of subsection 2.3 is 
constant in time for any Killing vector $K^a$ of the spacetime. 
Nevertheless, while for the conservation of $Q_m[K^a]$ built from 
the matter field variables $K^a$ must be a genuine Killing vector, 
the conservation of the analogous quantity $Q[K^a]$ of the 
matter+gravity system is ensured even by the {\it asymptotic} 
Killing fields too. 

However, if the time evolution is defined by the allowed time axis 
$\xi^a$ but $K^a$ is an asymptotic Killing vector with respect to 
another $\bar\xi^a$, i.e. $K^a\in{\cal A}^K_{\bar\xi}$, then in 
general $Q[K^a]$ is not conserved. In particular, if $\xi^a$ 
represents pure time translation at infinity (i.e. its lapse part 
tends to 1), but $K^a\in{\cal A}^K_{\bar\xi}$ for some $\bar\xi^a
\in{\cal G}$ (whenever $B_{\bi}(t)$, $R_{\bi\bj}(t)$, $T_{\bi}(t)$ 
and $T(t)$ of $K^a$ are independent of $t$), then $Q[K^a]$ is {\it 
not} constant in time with respect to $\xi^a$. For example, the 
time derivative of the centre-of-mass of Beig and \'O Murchadha 
with respect to $\xi^a$ above is not zero, that is just the 
spatial (linear) momentum. 

\bigskip
\bigskip

\ni
{\bf 4.2 The background-independence of $Q[K^a]$}\par
\bigskip
\ni
Let ${}_0v^a$ be the outward directed ${}_0q_{ab}$-unit normal to the 
coordinate spheres in $\Sigma$, and let $\gamma$ be an integral curve 
of ${}_0v^a$ form some ${\cal S}_{r_0}$ to ${\cal S}_r$. Then the 
length of $\gamma$ in the physical metric $q_{ab}$ is 

$$\eqalign{
R=\int^r_{r_0}\sqrt{\vert q_{ab}\,{}_0v^a\,{}_0v^b\vert}{\rm d}r'=&
 \int^r_{r_0}\sqrt{1-{1\over r'^k}q^{(k)}_{ab}\,{}_0v^a\,{}_0v^b+
 o(r'^{-k})}{\rm d}r'=\cr
=&\cases{r-r_0+A\ln{r\over r_0}+B+o(r^{-0}), &if $k=1$;\cr
 r-r_0+\bar A\,r^{-k+1}+\bar B+o(r^{-k+1}), &if $k\not=1$\cr} \cr}
\eqno(4.2.1)
$$
\ni
for some constants $A$, $\bar A$, $B$ and $\bar B$. This implies, in 
particular, that 

$$
{1\over R^k}-{1\over r^k}=\cases{ o(r^{-1}), &if $k=1$;\cr
             O(r^{-2k}), &if $k\not=1$.\cr}\eqno(4.2.2)
$$
\ni
Therefore, in the definitions (2.3.1)-(2.3.3) of the asymptotic 
flatness the radial distance $r$ can be substituted by the physical 
radial distance $R$ without changing the structure or the leading 
terms of $q_{\bi\bj}$ and $\chi_{\bi\bj}$. 

To clarify the potential ambiguity both of the notion of asymptotic 
flatness and the quantities $Q[K^a]$ coming from the non-uniqueness 
of the background metric ${}_0q_{ab}$, let $(\Sigma,q_{ab},\chi_{ab})$ 
be $(k,l)$-asymptotically flat with respect to ${}_0q_{ab}$, and let 
${}_0\bar q_{ab}$ be another background metric, being flat on $\Sigma
-K$. (Without loss of generality we may assume that the domain of the 
flatness of both ${}_0q_{ab}$ and ${}_0\bar q_{ab}$ coincide.) Thus 
there exists a diffeomorphism $\phi:\Sigma-K\rightarrow\Sigma-K$ such 
that ${}_0\bar q_{ab}=\phi^*\,{}_0q_{ab}$. For the sake of simplicity 
suppose that $\phi$ is homotopic to the identity ${\rm Id}\vert
_{\Sigma-K}$, i.e. for some one-parameter family $\phi_u$ of 
diffeomorphisms $\phi_0={\rm Id}\vert_{\Sigma-K}$ and $\phi_1=\phi$. 
Then we can form the one-parameter family of flat metrics ${}_0q_{ab}
(u):=\phi^*_u\,{}_0q_{ab}$ on $\Sigma-K$ (which can obviously be 
extended to the whole $\Sigma$ as, in general curved, negative 
definite metrics). If $V^a$ is the vector field on $\Sigma-K$ 
generating $\phi_u$, and its components in the coordinates $\{x
^{\bk}\}$ are defined by $V^a{}_0q_{ab}=:V_{\bk}{}_0D_bx^{\bk}$, then 

$$
\delta\,{}_0q_{ab}:=\bigl({{\rm d}\over{\rm d}u}{}_0q_{ab}(u)\bigr)
\vert_{u=0}=\L_{\bf V}{}_0q_{ab}=\Bigl({}_0D_{\bi}V_{\bj}+{}_0D_{\bj}
V_{\bi}\Bigr){}_0D_ax^{\bi}{}_0D_bx^{\bj}. \eqno(4.2.1)
$$
\ni
Writing $V_{\bi}$ in the form 

$$
V_{\bi}(x^{\bk})=2x^{\bk}\rho_{\bk\bi}+\tau_{\bi}+r^RV^{(R)}_{\bi}
\bigl({x^{\bk}\over r}\bigr)+o^\infty\bigl(r^R\bigr) \eqno(4.2.2)
$$
\ni
for some power $R$, where the first two terms together is just the 
kernel of the flat Killing operator ${}_0D_{({\bi}}V_{{\bj})}$ for 
${}_0q_{ab}$, we have 

$$
{}_0D_{({\bi}}V_{{\bj})}=r^{R-1}\Bigl(R\,{}_0v^{{\bk}}\delta_{{\bk}(
{\bi}}V^{(R)}_{{\bj})}+\bigl(\bar\partial_{\bk}V^{(R)}_{({\bi}}\bigr)
\bigl(\delta^{\bk}_{{\bj})}-\delta_{{\bj}){\bm}}\,{}_0v^{\bm}\,{}_0v
^{\bk}\bigr)\Bigr)+o^\infty\bigl(r^{R-1}\bigr).\eqno(4.2.3)
$$
\ni
Its leading term has even parity iff $V^{(R)}_{\bi}({x^{\bk}\over r})$ 
has odd parity. Since, for sufficiently small $u$, one has $q_{ab}-
{}_0q_{ab}(u)=q_{ab}-{}_0q_{ab}-({}_0q_{ab}(u)-{}_0q_{ab})=q_{ab}-{}_0
q_{ab}-\delta\,{}_0q_{ab}\,u+O(u^2)$, and hence, in the ${}_0q
_{ab}$-Cartesian coordinate system $\{x^{\bk}\}$, $q_{\bi\bj}-\,{}_0
q_{\bi\bj}(u)=q_{\bi\bj}-\,{}_0q_{\bi\bj}-\delta\,{}_0q_{\bi\bj}\,u+
O(u^2)=r^{-k}q^{(k)}_{\bi\bj}+o^\infty(r^{-k})-2\,{}_0D_{({\bi}}V
_{{\bj})}\,u+O(u^2)$ holds. The one-parameter family of coordinate 
systems $\{\bar x^{\bi}(u)\}$, defined in terms of the coordinates 
$\{x^{\bk}\}$ by $\bar x^{\bi}(u):=\phi^{\bi}_u(x^{\bk})$, is 
Cartesian with respect to the one-parameter family of flat metrics 
${}_0q_{ab}(u)$, i.e. ${}_0\bar q_{\bm\bn}(u)=-\delta_{\bm\bn}$. (To 
see this it is enough to recall the definition of the pull-back 
${}_0q_{ab}(u)$ of the metric 
${}_0q_{ab}$ along $\phi_u$ in the coordinates $\{x^{\bi}\}$, viz. 
${}_0q_{\bi\bj}(u):={\partial\phi^{\bm}\over\partial x^{\bi}}
{\partial\phi^{\bn}\over\partial x^{\bj}}\,{}_0q_{\bm\bn}=-\delta
_{\bm\bn}{\partial\bar x^{\bm}\over\partial x^{\bi}}{\partial\bar x
^{\bn}\over\partial x^{\bj}}$, and to compare this with the 
transformation law ${}_0q_{\bi\bj}(u)={}_0\bar q_{\bm\bn}(u){\partial
\bar x^{\bm}\over\partial x^{\bi}}{\partial\bar x^{\bn}\over\partial 
x^{\bj}}$ of the components ${}_0q_{\bi\bj}(u)$ and ${}_0\bar q_{\bm
\bn}(u)$ of ${}_0q_{ab}(u)$ in the coordinates $\{x^{\bi}\}$ and 
$\{\bar x^{\bm}(u)\}$, respectively.) 
Then for sufficiently small $u$ we have $\bar x^{\bi}=x^{\bi}+V^{\bi}
(x^{\bk})\,u+O(u^2)$. Thus the components of the physical metric and 
the extrinsic curvature in the ${}_0q_{ab}(u)$-Cartesian coordinate 
system $\{\bar x^{\bk}(u)\}$, respectively, are 

$$\eqalign{
\bar q_{\bi\bj}=&q_{\bm\bn}{\partial x^{\bm}\over\partial\bar x^{\bi}}
 {\partial x^{\bn}\over\partial\bar x^{\bj}}={}_0\bar q_{\bi\bj}+{1
 \over r^k}\Bigl(q^{(k)}_{\bi\bj}-2u\,q^{(k)}_{\bi\bk}\rho_{\bj}{}
 ^{\bk}-2u\,q^{(k)}_{\bj\bk}\rho_{\bi}{}^{\bk}\Bigr)+o^\infty\bigl(
 r^{-k}\bigr)-\cr
&-2u\,r^{R-1}\Bigl(R\,{}_0v^{\bk}\delta_{{\bk}({\bi}}V^{(R)}_{{\bj})}
 +\bar\partial_{({\bi}}V^{(R)}_{{\bj})}-\,{}_0v^{\bk}\bigl(\bar\partial
 _{\bk}V^{(R)}_{({\bi}}\bigr)\delta_{{\bj}){\bl}}\,{}_0v^{\bl}\Bigr)+
 o^\infty\bigl(r^{R-1}\bigr)+O\bigl(u^2\bigr),\cr
\bar\chi_{\bi\bj}=&\chi_{\bm\bn}{\partial x^{\bm}\over\partial\bar x
 ^{\bi}}{\partial x^{\bn}\over\partial\bar x^{\bj}}={1\over r^l}
 \Bigl(\chi^{(l)}_{\bi\bj}-2u\,\chi^{(l)}_{\bi\bk}\rho_{\bj}{}^{\bk}-
 2u\,\chi^{(l)}_{\bj\bk}\rho_{\bi}{}^{\bk}\Bigr)+o^\infty\bigl(r^{-l}
 \bigr)-\cr
&-2u\,r^{R-1-l}\Bigl(R\,{}_0v^{\bm}\delta_{{\bm}({\bi}}\chi^{(l)}
 _{{\bj}){\bk}}V^{(R){\bk}}+\chi^{(l)}_{{\bk}({\bi}}\bar\partial
 _{{\bj})}V^{(R){\bk}}-\,{}_0v^{\bm}\bigl(\bar\partial_{\bm}V^{(R)
 {\bk}}\bigr)\chi^{(l)}_{{\bk}({\bi}}\delta_{{\bj}){\bl}}\,{}_0v^{\bl}
 \Bigr)+\cr
&+o^\infty\bigl(r^{R-1-l}\bigr)+O\bigl(u^2\bigr).\cr}\eqno(4.2.4)
$$
\ni
Therefore, {\it the asymptotic end $(\Sigma,q_{ab},\chi_{ab})$, which 
is $(k,l)$-asymptotically flat with respect to the background 
metric ${}_0q_{ab}$, remains $(k,l)$-asymptotically flat with 
respect to the new flat metrics ${}_0q_{ab}(u)=\phi^*_u\,{}_0q_{ab}$ 
for sufficiently small $u$ precisely when $R\leq(1-k)$ and $V^{(R)}
_{\bi}$ has odd parity for $R=1-k$}. These changes of the background 
metrics will be called {\it allowed}. (Apart from a rigid Euclidean 
rotation, the non-trivial leading terms of $q_{\bi\bj}$ and $\chi
_{\bi\bj}$ are invariant with respect to the change ${}_0q_{ab}\mapsto
\,{}_0q_{ab}(u)$ of the background metric iff $R<(1-k)$.) Therefore, 
the corresponding $V^a$'s are special, purely spatial asymptotic 
Killing vectors. 

The change of the background metric ${}_0q_{ab}$ yields a change of 
the allowed time axes. By (4.2.2) for the infinitesimal change of 
$M$ and $M_a$ given by (2.4.1) one has 

$$\eqalign{
\delta M:=&\Bigl({{\rm d}\over{\rm d}u}M\bigl(\bar x^{\bk}(u)\bigr)
 \Bigr)_{u=0}=\cr
=&2V^{\bi}B_{\bi}+r^{E-1}\Bigl(E\,{}_0v^{\bk}\delta_{\bk\bi}\nu^{(E)}
 +\bigl(\bar\partial_{\bk}\nu^{(E)}\bigr)\bigl(\delta^{\bk}_{\bi}-{}_0
 v^{\bk}\,{}_0v^{\bj}\delta_{\bj\bi}\bigr)\Bigr)V^{\bi}+o^\infty\bigl(
 r^E\bigr)=\cr
=&2x^{\bk}\bigl(2\rho_{\bk\bj}B^{\bj}\bigr)+2\tau_{\bj}B^{\bj}+2r^E\,
 {}_0v^{\bk}\rho_{\bk}{}^{\bi}\bigl(\bar\partial_{\bi}\nu^{(E)}\bigr)+
 2r^RB^{\bi}V^{(R)}_{\bi}+o^\infty\bigl(r^E\bigr)+o^\infty\bigl(r^R
 \bigr),\cr
\delta M_a:=&\Bigl({{\rm d}\over{\rm d}u}\bigl(M_{\bi}\bigl(\bar x
 ^{\bk}(u)\bigr)\,{}_0D_a\bar x^{\bi}(u)\bigr)\Bigr)_{u=0}=\cr
=&\Bigl(2x^{\bk}\bigl(2\rho_{\bk}{}^{\bj}R_{\bj\bi}-2R_{\bk}{}^{\bj}
 \rho_{\bj\bi}\bigr)+\bigl(2\tau^{\bj}R_{\bj\bi}-2T^{\bj}\rho_{\bj\bi}
 \bigr)+2r^F\bigl(\,{}_0v^{\bk}\rho_{\bk}{}^{\bj}\bigl(\bar\partial
 _{\bj}\nu^{(F)}_{\bi}\bigr)-\nu^{(F)}_{\bj}\rho^{\bj}{}_{\bi}\bigr)+\cr
+&2r^R\,{}_0v^{\bk}R_{\bk}{}^{\bj}\bigl(R\,{}_0v^{\bm}\delta_{\bm\bi}
 V^{(R)}_{\bj}+\bigl(\bar\partial_{\bm}V^{(R)}_{\bj}\bigr)\bigl(\delta
 ^{\bm}_{\bi}-{}_0v^{\bm}\,{}_0v^{\bn}\delta_{\bn\bi}\bigr)\bigr)+o
 ^\infty\bigl(r^F\bigr)+o^\infty\bigl(r^R\bigr)\Bigr)\,{}_0D_ax^{\bi}. 
 \cr}\eqno(4.2.5)
$$
\ni
Thus the allowed change of the background metric acts on $K^a=Mt^a
+M^a$ as the diffeomorphism generated by the asymptotic Killing 
vector $V^a$: for $E,F\leq(1-k)$ the structure of $\delta M$ and 
$\delta M_a$ is similar to that of $M$ and $M_a$, respectively, if 
$R<(1-k)$ or if $R=(1-k)$ and $V^{(R)}_{\bi}$ has odd parity. If, 
however, $B_{\bi}=0$, $R_{\bi\bj}=0$ and $E,F\leq-k$, i.e. $K^a\in{}
_{-k}{\cal T}^K_\xi$, then the structure of $\delta M$ and $\delta 
M_a$ will be similar to that of $M$ and $M_a$, respectively, only 
if $R\leq-k$. 

To calculate the change of $Q[K^a]$ under the allowed change of the 
background metric let us form $Q_u[K^a(u)]$ by using the one-parameter 
family of the flat background metrics ${}_0q_{ab}(u)$ in (4.1.1) 
instead of ${}_0q_{ab}$. Then, with the notation $\delta K^a:=\delta M
t^a+q^{ab}\delta M_b$, a straightforward calculation gives 

$$\eqalign{
\delta Q[K^a]:&=\bigl({{\rm d}\over{\rm d}u}Q_u[K^a(u)]\bigr)\vert
 _{u=0}=\cr
&=Q[\delta K^a]+{1\over2\kappa}\int_\Sigma D_a\Bigl\{M\bigl(\delta\,{}_0
 \Gamma^a_{bc}q^{bc}-\delta\,{}_0\Gamma^c_{cb}q^{ba}\bigr)+\bigl(q
 ^{ab}q^{cd}-q^{ac}q^{bd}\bigr)\bigl(\,{}_0D_bM\bigr)\delta\,{}_0
 q_{cd}\Bigr\}{\rm d}\Sigma,\cr}\eqno(4.2.6)
$$
\ni
where $\delta\,{}_0\Gamma^a_{bc}:={1\over2}\,{}_0q^{ad}(-{}_0D_d\delta
\,{}_0q_{bc}+{}_0D_c\delta\,{}_0q_{db}+{}_0D_b\delta\,{}_0q_{cd})$. 
Thus the spatial momentum and angular momentum depend on the 
background metric only through their generator $M_a$, but the 
energy and centre-of-mass may be ambiguous as a consequence of 
the non-vanishing of the integral on the right of (4.2.6) too. 
(Using the expression (4.1.2) for $Q[K^a]$ instead of (4.1.1) it 
can be shown that $\delta Q[K^a]$ is finite if $R<(1-k)$, or if 
$R=(1-k)$ and $V^{(R)}_{\bi}$ has odd parity.) First suppose that 
$K^a=Mt^a+M^a$ is an asymptotic Killing vector, and hence $E,F
\leq(1-k)$. Then by the results of the previous subsection $k\geq
{1\over2}(n-1)$ must hold, and by (4.2.4) and (4.2.5) $R\leq(1-k)$ 
must be required. Thus $Q[\delta K^a]$ depends only on the first 
two terms of $\delta M$ and $\delta M_a$ in (4.2.5), because the 
remaining terms are pure gauge generators. However, by (2.4.11-14) 
this is nothing but the transformation law of the energy, and the 
components of the spatial momentum, spatial angular momentum and 
centre-of-mass under the Euclidean transformation coming from the 
diffeomorphism generated by $V^a$. To rule out the ambiguities in 
the expression of the energy and the centre-of-mass, we must 
require the vanishing of the integral in (4.2.6). Its vanishing 
can be ensured by requiring $R<(2-n)$, or $R=(2-n)$ and that $V^{(R)}
_{\bi}$ be odd parity functions. Next suppose that $K^a\in{}_{-k}
{\cal T}^K_\xi$, i.e. $R_{\bi\bj}=0$, $B_{\bi}=0$ and $E,F\leq-k$. 
Then by the previous subsection $Q[K^a]$ is finite even if $k>{1
\over2}(n-2)$, and by (4.2.5) $R\leq-k$, implying that $Q[\delta 
K^a]$ is finite and describes how the components of the spatial 
momentum transform under the Euclidean transformation coming from 
the diffeomorphism generated by $V^a$. (The energy remains intact.) 
The vanishing of the integral in (4.2.6) can be ensured even by $R
<(3-n)$. 
Therefore, {\it $Q[K^a]$ is unambiguously defined for $K^a\in{\cal 
A}^K_\xi$ if $R\leq(1-k)$, $R\leq(2-n)$ and in the case of the 
equality, $R=(2-n)$, $V^{(R)}_{\bi}$ has odd parity; and for $K^a
\in{}_{-k}{\cal T}^K_\xi$ if $R\leq-k$ and $R<(3-n)$}. In particular, 
in $m=3+1$ spacetime dimensions and for $k=1$ the angular momentum 
and centre-of-mass are well defined provided the diffeomorphisms 
connecting the background metrics tend to the rigid Euclidean 
transformations like $O(r^{-1})$ with odd parity generator, or 
faster. For the condition ensuring well defined energy and spatial 
momentum we recovered the known results of [8,11]. Namely, writing 
$k$ in the form $k={1\over2}+\delta$ for some $\delta>0$, the energy 
and momentum are well defined if the diffeomorphisms tend to the 
Euclidean transformation like $O(r^{-{1\over2}-\delta})$. 

\bigskip
\bigskip
\ni
{\lbf Appendix: Global integral conditions at the null infinity}\par
\medskip
\ni
Let us consider global quantities associated to the global state of 
the matter fields in the Minkowski spacetime at a given {\it retarded 
time U}, i.e. if the spacelike hypersurface of subsection 2.2 extends 
to the {\it future null infinity} ${\cal I}^+$ such that its 
intersection with ${\cal I}^+$ is the $U={\rm constant}$ cut. However, 
in the present context it seems more comfortable to choose this 
hypersurface to be the null hypersurface ${\cal N}_U:=\{(\tau,X^{\bi})
\vert\,\,\tau-R=U\,\}$ instead of a spacelike (e.g. a hyperboloidal) 
one. Here $(\tau,X^{\bi})$, ${\bi}=1,...,n$, are still the Cartesian 
coordinates introduced in subsection 2.2. For the null ${\cal N}_U$ 
the flux integral of $T^{ab}K_b$ takes the form $\lim_{R\rightarrow
\infty}\int_0^R\oint_{\cal S}T^{ab}l_aK_b{\rm d}{\cal S}R'^{n-1}{\rm 
d}R'$, where $l_a:=\nabla_a(\tau-R)=t_a+v_a$, a null normal to ${\cal 
N}_U$. Thus if, for the sake of brevity, we define $\rho(U,R,{X^{\bk}
\over R}):=t_aT^{ab}l_b=\mu(U+R,X^{\bk})+j^a(U+R,X^{\bk})v_a$ and 
$\rho^{\bi}(U,R,{X^{\bk}\over R}):=K^{\bi}_aT^{ab}l_b=K^{\bi}_a(j^a(U+
R,X^{\bk})+\sigma^{ab}(U+R,X^{\bk})v_b)$, then the global energy, 
spatial momentum, angular momentum and centre-of-mass measured at the 
retarded time $U$ at future null infinity are finite precisely when 
$\oint_{\cal S}{}^+\rho{\rm d}{\cal S}=o(R^{-n})$, $\oint_{\cal S}{}^+
\rho^{\bi}{\rm d}{\cal S}=o(R^{-n})$, $\oint_{\cal S}{}^-\rho^{[{\bi}}
X^{{\bj}]}{1\over R}{\rm d}{\cal S}=o(R^{-m})$ and $\oint_{\cal S}({}
^+\rho^{\bi}-{}^-\rho{X^{\bi}\over R}){\rm d}{\cal S}=o(R^{-m})$, 
respectively. These conditions could be satisfied if $\rho(U,R,{X
^{\bk}\over R})={1\over R^m}\rho^{(m)}(U,{X^{\bk}\over R})+o(R^{-m})$ 
and $\rho^{\bi}(U,R,{X^{\bk}\over R})={X^{\bi}\over R}({}^+\varphi(U,
R,{X^{\bk}\over R})+{1\over R^m}{}^-\rho^{(m)}(U,{X^{\bk}\over R}))+
o(R^{-m})$, where ${}^-\rho^{(m)}(U,{X^{\bk}\over R})$ is the odd 
parity part of $\rho^{(m)}(U,{X^{\bk}\over R})$ and ${}^+\varphi(U,
R,{X^{\bk}\over R})$ is an arbitrary function with even parity. ${}
^\pm\rho^{(m)}$ contribute to the energy and spatial momentum, but 
do not to the angular momentum and centre-of-mass. Thus we may 
call them the {\it BS mass aspect} of $T^{ab}$. 

\bigskip
\bigskip

\ni
{\lbf Acknowledgment}\par
\medskip
\ni
The author is grateful to Robert Beig, Jacek Jezierski, James 
Nester and Alexander Petrov for their valuable remarks on angular 
momentum at spatial infinity. Special thanks to Alexander Petrov 
for his important remarks on the preliminary version of the present 
paper. This work was partially supported by the Hungarian Scientific 
Research Fund grants OTKA T030374 and T042531. 

\bigskip
\bigskip
\ni
{\lbf References}\par
\bigskip
\item{[1]} R. Haag, D. Kastler, An algebraic approach to quantum 
           field theory, J. Math. Phys. {\bf 5} 848-861 (1964)
\item{   } R. Haag, {\it Local Quantum Physics, Fields, Particles, 
           Algebras}, Springer Verlag, Berlin and Heidelberg 1992

\item{[2]} J.D. Brown, J.W. York, Quasi-local energy and conserved 
           charges derived from the gravitational action, Phys. Rev. 
           D {\bf 47} 1407-1419 (1993) 

\item{[3]} L.B. Szabados, Quasi-local energy-momentum and two-surface
           characterization of the pp-wave spacetimes, Class. Quantum 
           Grav. {\bf 13} 1661-1678 (1996)
\item{   } L.B. Szabados, On certain quasi-local spin-angular 
           momentum expressions for large spheres near the null 
           infinity, Class. Quantum Grav. {\bf 18} 5487-5510 (2001), 
           Corr.: Class. Quantum Grav. {\bf 19} 2333 (2002)

\item{[4]} T. Regge, C. Teitelboim, Role of surface integrals in the 
           Hamiltonian formulation of general relativity, Ann. Phys. 
           {\bf 88} 286-318 (1974)
\item{[5]} R. Beig, N.\'O Murchadha, The Poincar\'e group as the 
           symmetry group of canonical general relativity, Ann. Phys. 
           {\bf 174} 463-498 (1987)

\item{[6]} D. Baskaran, S.R. Lau, A.N. Petrov, Center of mass integral 
           in canonical general relativity, gr-qc/ 0301069 

\item{[7]} J.M. Nester, The gravitational Hamiltonian, in {\it 
           Asymptotic behavior of mass and spacetime geometry}, Ed. 
           F.J. Flaherty, Lecture Notes in Physics No 202, pp. 
           155-163, Springer 1984 

\item{[8]} P.T. Chru\'sciel, Boundary conditions at spacelike infinity 
           from a Hamiltonian point of view, in {\it Topological and 
           global structure of spacetime}, NATO Adv. Sci. Inst. Ser. 
           B Phys. No. 138, pp. 49-59, Plenum Press, New York 1986 
\item{[9]} R. Bartnik, The mass of an asymptotically flat manifold, 
           Commun. Pure Appl. Math. {\bf 39} 661-693 (1986)
\item{[10]} N. \'O Murchadha, Total energy-momentum in general 
           relativity, J. Math. Phys. {\bf 27} 2111-2128 (1986)
\item{[11]} P.T. Chru\'sciel, A remark on the positive-energy theorem, 
            Class. Quantum Grav. {\bf 3} L115-L121 (1986)
\item{[12]} P. Bizo\'n, E. Malec, On Witten's positive-energy proof for 
            weakly asymptotically flat spacetimes, Class. Quantum Grav. 
            {\bf 3} L123-L128 (1986)

\item{[13]} R. Geroch, {\it Asymptotic structure of spacetime}, in 
            Asymptotic Structure of Spacetime, ed. F.P. Esposito and 
            L. Witten, Plenum Press, New York 1977
\item{[14]} A. Ashtekar, R.O. Hansen, A unified treatment of null 
            and spatial infinity in general relativity. I. Universal 
            structure, asymptotic symmetries and conserved quantities 
            at spatial infinity, J. Math. Phys. {\bf 19} 1542-1566 
            (1978)
\item{[15]} R. Beig, B.G. Schmidt, Einstein's equations near spatial 
            infinity, Commun. Math. Phys. {\bf 87} 65-80 (1982) 
\item{[16]} H. Friedrich, Gravitational fields near space-like and 
            null infinity, J. Geom. Phys. {\bf 24} 83-163 (1998) 

\item{[17]} R.M. Wald, A. Zoupas, A general definition of ``conserved 
            quantities'' in general relativity and other theories of 
            gravity, Phys. Rev. D {\bf 61} 084027-1-16 (2000) 
\item{[18]} A. Ashtekar, S. Das, Asymptotically anti-de-Sitter 
            spacetimes: Conserved quantities, Class. Quantum Grav. 
            {\bf 17} L17-L30 (2000) 

\item{[19]} N.M.J. Woodhouse, {\it Geometric quantization}, Clarendon 
            Press, 1980; revised edition in 1991
\item{[20]} A. Ashtekar, {\it New perspectives in canonical gravity}, 
            Bibliopolis, Naples 1988
\item{[21]} R.M. Wald, {\it General relativity}, University of Chicago 
            Press, 1984
\item{[22]} J. Lee, R.M. Wald, Local symmetries and constraints, J. 
            Math. Phys. {\bf 31} 725-743 (1990)

\item{[23]} L.B. Szabados, On the role of conformal three-geometries 
            in the dynamics of general relativity, Class. Quantum 
            Grav. {\bf 19} 2375-2391 (2002)
\item{[24]} S.W. Hawking, G.F.R. Ellis, {\it The large scale structure 
            of spacetime}, Cambridge Univ. Press, Cambridge 1973
\item{[25]} R. Beig, N.\'O Murchadha, The momentum constraint of 
            general relativity, Commun. Math. Phys. {\bf 176} 723-738 
            (1996)
\item{[26]} L.B. Szabados, On certain global conformal invariants 
            and 3-surface twistors of initial data sets, Class. 
            Quantum Grav. {\bf 17} 793-811 (2000)

\item{[27]} R. Beig, The classical theory of canonical general 
            relativity, in {\it Canonical Gravity: From Classical to 
            Quantum}, Springer Lecture Notes in Physics No 434, 
            Ed. J. Ehlers, H. Friedrich, pp. 59-80, Springer--Verlag, 
            1994
\item{[28]} V.O. Solovev, Generator algebra of the asymptotic 
            Poincare group in the general theory of relativity, Theor. 
            Math. Phys. {\bf 65} 1240-1249 (1985) 
\item{[29]} A.N. Petrov, Asymptotically flat spacetimes at spatial 
           infinity: The field approach and the Lagrangian 
           description, Int. J. Mod. Phys. D {\bf 4} 451-478 
           (1995)
\item{[30]} A.N. Petrov, Asymptotically flat spacetimes at spatial 
           infinity : II Gauge invariance of the integrals of 
           motion in the field approach, Int. J. Mod. Phys. D 
           {\bf 6} 239-261 (1997)

\end